\documentclass[useAMS,usenatbib]{mn2e}
\usepackage{graphicx}
\usepackage{amssymb}
\usepackage{rotating}

\newcommand{\apj}{ApJ}

\newcommand{\aj}{AJ}
\newcommand{\mnras}{MNRAS}

\newcommand{\MC}{\multicolumn}
\newcommand{\kms}{km~s$^{-1}$}

\newcommand{\HI}{H{\sc i}}
\newcommand{\HII}{H{\sc ii}}
\newcommand{\sunn}{$_{\odot}$}

\title[LV dwarf KK242]
{Local Volume dwarf KK242: radial velocity, SF region, and metallicity}
\author[S.A.~Pustilnik et al.]
{S.A.~Pustilnik,$^{1}$\thanks{E-mail: sap@sao.ru (SAP)}
A.L.~Tepliakova,$^{1}$ Y.A.~Perepelitsyna,$^{1}$ A.Y.~Kniazev,$^{2,3,4,1}$
\newauthor L.N.~Makarova,$^{1}$ A.N.~Burenkov,$^{1}$ S.S.~Kotov,$^{1}$  E.A.~Malygin$^{1}$ \\
$^1$ Special Astrophysical Observatory of RAS, Nizhnij Arkhyz,
Karachai-Circassia 369167, Russia\\
$^2$ South African Astronomical Observatory, PO Box 9, 7935 Observatory,  Cape Town, South Africa \\
$^3$ Southern African Large Telescope Foundation, PO Box 9, 7935 Observatory, Cape Town, South Africa \\
$^4$ Sternberg Astronomical Institute, Lomonosov Moscow State University, Universitetskij Pr. 13,  Moscow 119992, Russia
}

\begin{document}

\label{firstpage}

\date{Accepted 2022 Spetember 9. Received 2022 September 5; in original form 2022 June 21}

\pagerange{\pageref{firstpage}--\pageref{lastpage}} \pubyear{2022}

\maketitle

\begin{abstract} 
KK242 is a LV dwarf of transition type residing in the void environment.
Koda et al. present clear indications on its connection with
Scd galaxy NGC6503. This implies the distance to KK242 of $\sim$6.3~Mpc
and its M$_{\rm B}$ = --10.5~mag.
Its radial velocity, known from the Effelsberg radio telescope \HI\
observations, reveals,
however, the difference with that of NGC6503, $\Delta$V $\sim$ 400~\kms.
If real, this fact implies the substantial constraints on its origin. To
clear-up the issue of KK242 radial velocity, we obtained with the SAO 6-m telescope
spectra of its faint star-forming (SF) complex.
H$\alpha$ and H$\beta$ emission is detected in two adjacent compact
regions, the southern and northern, separated by $\sim$2~arcsec
($\sim$60 pc). Their mean radial velocity is V$_{\rm hel}$ = --66~\kms,
$\sim$100~\kms\ lower than that of NGC6503.
We use the HST Legacy Archive images and photometry of individual stars
from the Extragalactic Distance Database, available for KK242, to identify in the SF complex the
exciting hot stars, the probable BHeB and RHeB stars and a supernova remnant.
We address, based on the possible range of its gas metallicity, the probable
evolutionary paths of KK242. Using package {\sc Cloudy} and parameters of the exciting
B0V stars, we conclude that the observed flux ratio of [S{\sc ii}]
doublet to H$\alpha$   is consistent with the value of 12+log(O/H) $\sim$7.35 $\pm$ 0.18~dex,
expected for a stripped void dIrr galaxy.
\end{abstract}


\section[]{Introduction}
\label{sec:intro}
\setcounter{figure}{0}

A low surface brightness (LSB) dwarf galaxy of 'transition type' (dTr)
KK242 was first discovered by
\citet{KK1998} as a potential companion of Scd galaxy NGC6503.
\citet{HKKE2000} observed KK242 in the 21-cm \HI-line with the Effelsberg
100-m radio telescope and found an emission line at the radial velocity of
426~\kms. The latter differs from the radial velocity of NGC6503 by
$\sim$ 400~\kms\ that looks quite unusual. Later, \citet{Koda2015} rediscovered
this galaxy in the deep images of their survey for LSB dwarfs around spiral
galaxies. They presented images of KK242 (their name NGC6503-d1) in various
filters, including H$\alpha$.
Besides, they resolved in the broad-band images about three hundred the
brightest stars of KK242. From the analysis of the colour-magnitude diagram (CMD),
they derived the TRGB-based (Tip of Red Giant Branch) distance estimate
consistent with that known for NGC6503.

Both galaxies fall to the region of a nearby void occupying a part of the
Local Volume (hereafter LV,
see for more detail Section~\ref{ssec:environs}). In the framework of the
ongoing project aimed in studying various subsamples of void
galaxies in the Nearby Void Galaxy (NVG) catalog
\citep[see][]{PTM19, XMP-SALT, XMP-BTA}, we conduct, in particular, their
spectral observations to derive their gas metallicities and/or improve
the accuracy of radial velocities. As said above, \citet{Koda2015} show
that there is a compelling evidence of KK242 to be a companion of NGC6503.
However, its known estimate of the radial velocity, based on
a single \HI\ observation with the Effelsberg 100-m radio telescope, is too
much deviating from that of the host Scd galaxy.
This may 'provoke' various 'exotic' scenarios of the origin of this pair.
This was the primary motivation to clear up the issue of KK242 radial velocity.

For this end, we obtained the spectrum  of the only known faint
complex of H$\alpha$ emission in KK242 detected in papers of \citet{Koda2015}
and \citet{Kaisin2019}. The derived here radial velocity of --66~\kms\ differs
drastically from that measured via the single dish 21-cm \HI\ line emission.
Recently \citet{KK242.HST}
used our H$\alpha$ radial velocity to search for the possible \HI\ 21-cm line emission
based on the VLA D-configuration data cube for KK242. They detected the very faint
\HI\ line at the position of KK242 with the radial velocity of V(HI) = --80~\kms,
which is consistent within uncertainties with V(H$\alpha$) obtained in
this work.

The estimate of the radial velocity of KK242 via its H$\alpha$ line was our primary
task. Fortunately, thanks to the appropriate seeing during these observations and
a suitable position angle of the long slit, we also got the interesting
by-product results related to the substructure of this emission region
and its individual components. Coupled with the {\it Hubble Space Telescope
(HST)} Advanced Camera for Surveys (ACS) images from the Hubble Legacy
Archive (HLA) and with the photometry of
individual stars available in the Extragalactic Distance Database
(EDD) \citep[][and references therein]{Anand21}, this enables us
to discuss this star-forming complex in a more detail.

The galaxy KK242 is also interesting for a deeper insight as one
of a few known dTr objects in the void environment.
The great majority of the 30 known dTrs within the distances of
5~Mpc \citep{KK258} are related to a more typical
environment like the Local Group and similar nearby groups. Such a connection
can be related
to the origin of this type dwarfs. Only two of these 30 dTr, UGC1703 and
KK258 reside within the nearby voids described in \citet{PTM19}. Two more
dTrs, KKs03 and
DDO210, are well isolated, despite the latter is situated close to the border
of the Local Group.

In this context, KK242 as a representative of the small minority of
the void galaxy population,
might display various deviations from other known dwarfs of this rare type.
Besides, it is interesting to study the evolutionary path of such an unusual
dwarf in the void-type global environment. The gas metallicity, if it could be
estimated, is one of the important  parameters used for the comparison
of the observed properties of KK242 with those expected in the variety
of possible evolutionary scenarios. Moreover, the issue of the massive star
formation in such an atypically low-gas-density dwarf is crucial to address
since this case represents even a more extreme gas environment than in
the late-type gas-normal LSB dwarfs.

The rest of the paper is arranged as follows.
In Sec.~\ref{sec:obs}, the spectral observations and data reduction
are outlined and the used archive {\it HST} images are described.
In Sec.~\ref{sec:results}, the results of the analysis of the
BTA spectra along with the identification of the related objects
at the {\it HST} image are presented.
In Sec.~\ref{sec:discuss}, we discuss the properties of the studied SH
region in KK242, the range of metallicity and the global environment of KK242,
and in Sec.~\ref{sec:summary}, we summarize our results and conclude.
The linear scale at the adopted distance to KK242 (6.36 Mpc) is
30.9~pc in 1~arcsec.

\section[]{Observations and data processing}
\label{sec:obs}

\begin{table}
\centering{
\caption{{\bf Journal of BTA observations of KK242}}
\label{tab:journal}
\begin{tabular}{l|l|l|c|c} \\ 
\hline     
\MC{1}{c|}{Date} &
\MC{1}{c|}{Grism}&
\MC{1}{c|}{Expos.}&
\MC{1}{c|}{$\beta$}&
\MC{1}{c}{Air mass}  \\

\MC{1}{c|}{ } &
\MC{1}{c|}{} &
\MC{1}{c|}{time, s}&
\MC{1}{c|}{arcsec} &
\MC{1}{c}{}  \\

\hline     
2020.11.11 & VPHG1200R    & 3$\times$900  & 1.3 & 1.30 \\ 
2021.11.05 & VPHG1200@540 & 4$\times$1200  & 1.1 & 1.34 \\ 
2022.07.29 & VPHG1200R    & 8$\times$900  & 1.3 & 1.25 \\ 
\hline  
\end{tabular}
}
\end{table}

\begin{figure*}
\centering{
\includegraphics[width=9.0cm,angle=-0,clip=]{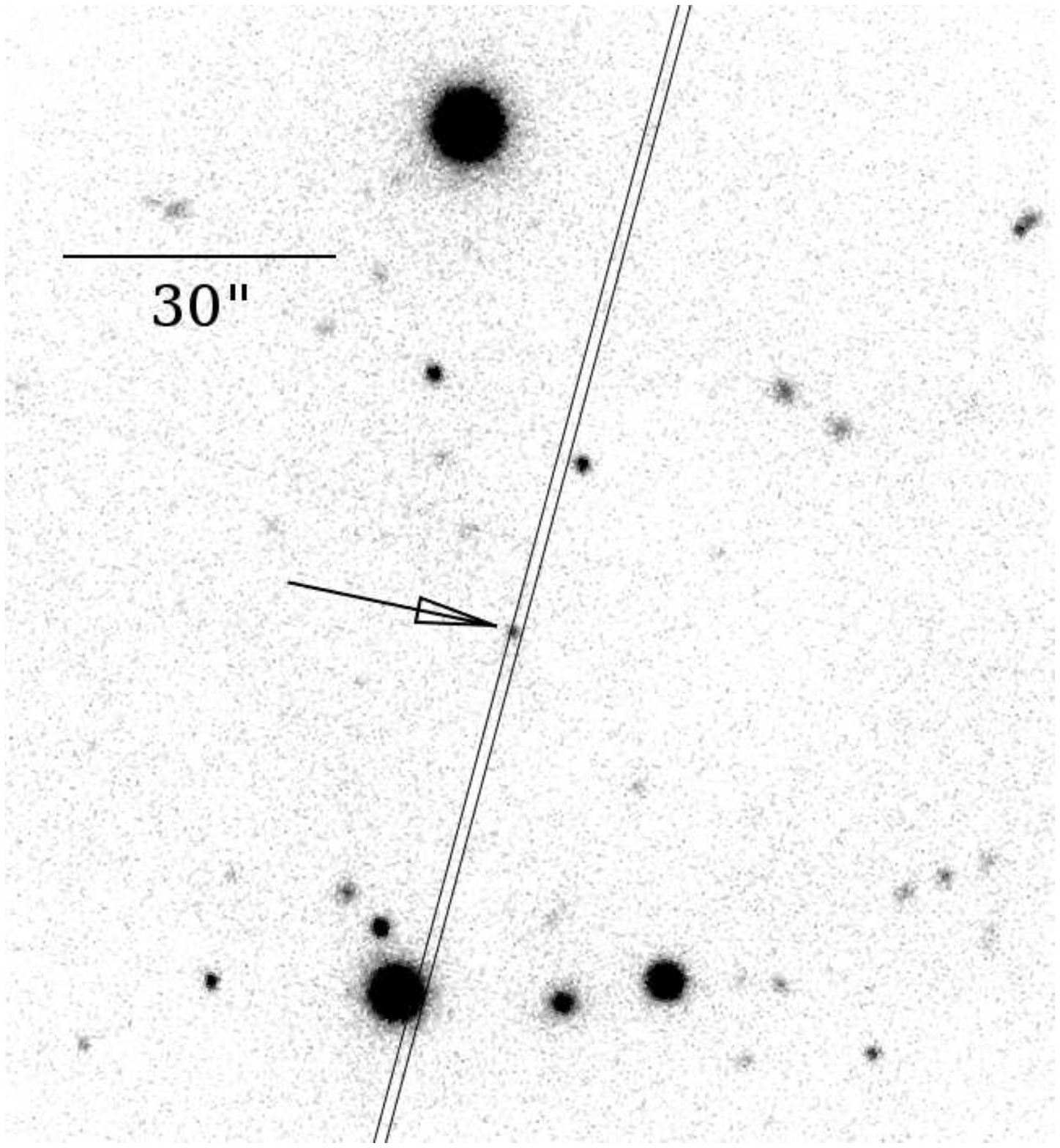}
\includegraphics[width=7.0cm,angle=-0,clip=]{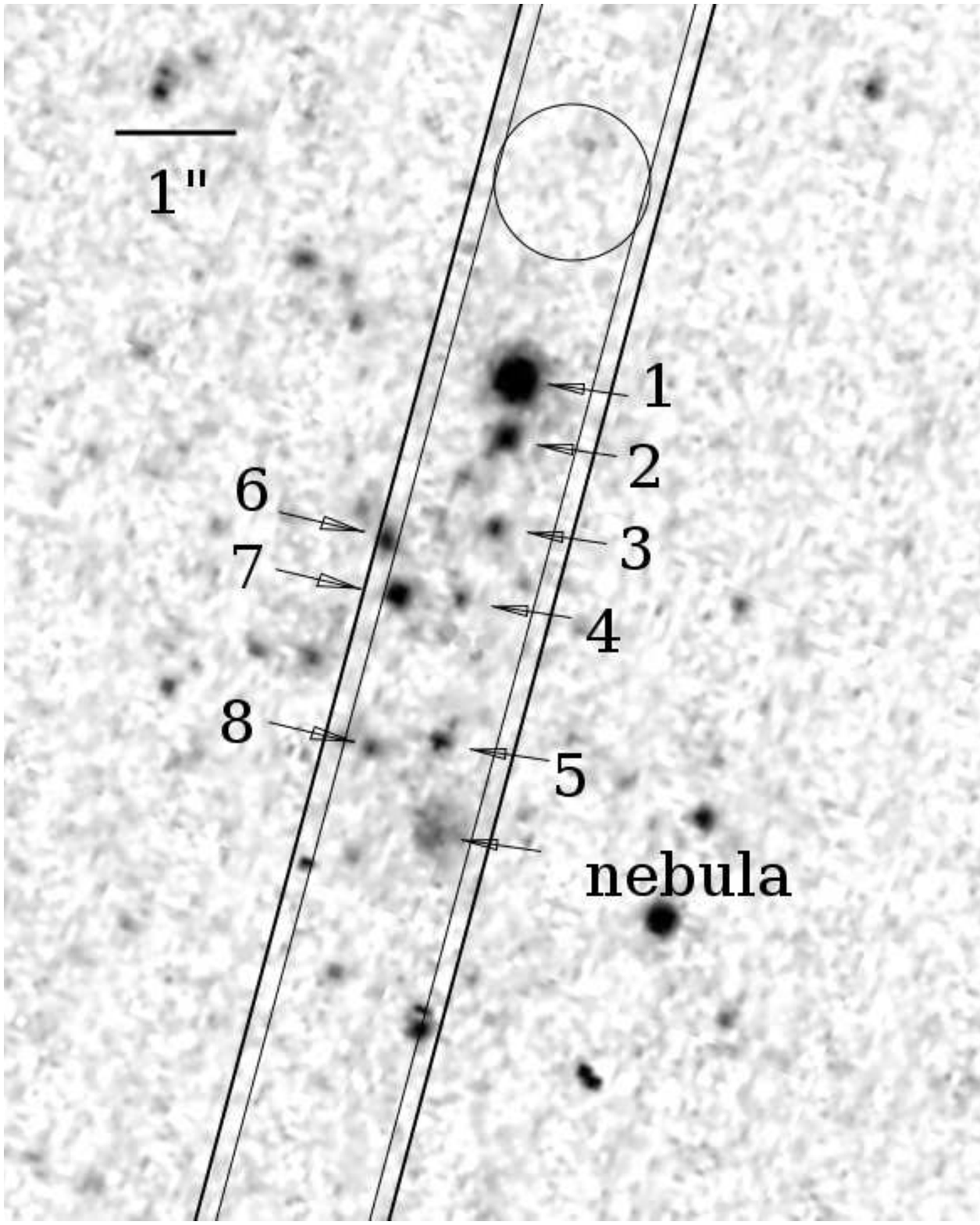}
\caption{Slit positions (PA = --15\degr) overlaid on the
BTA R-band ({\bf left panel}) and the HST F606W filter (ProjID 15922) ({\bf right panel})
images. N is up, E is to the left. Thin lines show SCORPIO-1 slit (width = 1.2~arcsec),
thick lines - SCORPIO-2 slit (width = 1.54~arcsec). A 'nebula' in the HST image, with
the apparent diameter of $\sim$0.5~arcsec, is at $\sim$ 3.3~arcsec to the South from
the brightest star of the frame (No.~1).
Both objects are visible on the BTA 2d spectrum in Fig.~\ref{fig:Spectra} (top
panel). Arrows show the brightest stars within the BTA slit. Their parameters are given
in Table~\ref{tab:HSTdata}. The scales, 30 and 1~arcsec, for left and right panels, respectively,
are shown by horizontal bars. Circle in the HST image shows the FWHM = 1.3~arcsec seeing
for the SCORPIO-1 spectrum.
 } }
\label{fig:HST_ima}
\end{figure*}

We obtained three optical spectra of KK242.
The first spectrum of KK242 was obtained with the BTA multimode
instrument SCORPIO-1 \citep{SCORPIO} during the night 2020 November 11,
under photometric conditions (see Table~\ref{tab:journal}). The
long slit with the width of 1.2~arcsec and the scale along the slit
of 0.36~arcsec pixel$^{-1}$ (after binning by 2) was positioned on the brightest {\it R}-band
source along the elongation of H$\alpha$ emission, corresponding to PA =
--15\degr (see the left-hand panel of Fig.~1).
For a more detailed description of the slit position relative to the
emission-line regions, see Sect.~\ref{ssec:HST}.
The grism VPHG1200R with the 2K$\times$2K CCD detector E2V~42-40
(13.5$\times$13.5~$\mu$m pixel) provided the spectrum coverage of 5700--7500~\AA\
with the FWHM $\sim$ 5.0~\AA.

The second spectrum of KK242 was obtained with the next generation BTA
multimode instrument SCORPIO-2 \citep{SCORPIO-2} during the night 2021 November 5,
under photometric conditions (see Table~\ref{tab:journal}). We aimed to
pick up in this spectrum all the light collected in the first observation.
Therefore, accounting for the smaller seeing on this night ($\sim$1.1 versus 1.3 arcsec),
we select of the two possible slit width options, 1.0 and 1.54 arcsec,
the wider one. This should allow us to directly compare the results for
both spectra.
The long slit with the width of 1.54~arcsec and the scale along the slit
of 0.40~arcsec pixel$^{-1}$ (after binning by 2) was positioned similar to that in the first
observations with PA = --15\degr. The grism VPHG1200@540 with
the 4K$\times$2K CCD detector E2V 261-84 (15$\times$15~$\mu$m pixel)
provided the spectrum coverage of 3650--7250~\AA\  with the FWHM $\sim$6.0~\AA.

The third spectrum of KK242 was obtained with SCORPIO-1 and grism VPHG1200R
 during the night 2022 July 29, with the similar set-up as for the first observation.
The long slit for this observation was positioned exactly as for the first time.
Since the seeing also was close to that of the first observation, the main
difference was the total integration time: 7200 sec in July 2022 versus 2700 sec
in November 2020. The main goal of the latter observation was to improve the S-to-N
ratio for [S{\sc ii}]$\lambda\lambda$6716,6731 doublet in the resulting average
spectrum, since from the  previous
data its uncertainty was too high to come to more or less confident conclusion on
the KK242 gas metallicity.

The main procedures of data reduction are described in
\citet{PaperVII}. Here we briefly outline them.
Our standard pipeline with the use of {\sc IRAF}\footnote{{\sc IRAF}: the Image
Reduction and Analysis Facility is distributed by the National Optical
Astronomy Observatory, which is operated by the Association of Universities
for Research in Astronomy, Inc. (AURA) under cooperative agreement with the
National Science Foundation (NSF).}
and {\sc MIDAS}\footnote{{\sc MIDAS} is an acronym for the European Southern
Observatory package -- Munich Image Data Analysis System. }
was applied for the reduction of long-slit spectra. It includes the
following steps: removal of cosmic ray hits, bias subtraction, flat-field
correction, wavelength calibration, night-sky background subtraction.
Spectrophotometric standard stars observed during these nights,
were used to obtain spectra in the absolute flux scale.

In the resulting 2d spectra of all observations of KK242,
two distinct H$\alpha$ knots are seen with centres separated along the slit
by $\sim$2~arcsec. In Fig.~3  (top panel), we show a part
of the 2d spectrum for the first night. The Southern knot is about twice
brighter in H$\alpha$, and has a much weaker underlying continuum in comparison
to that of the N knot.
The 1d spectra for the S and N knots were extracted, summing up 5 or 6 (for different
detectors) and respectively 6 or 7 pixels along
the slit ($\sim$ 2.1 and 2.5~arcsec, respectively), without weights, centred
on each of two maxima of the H$\alpha$ line signal.
We have no opportunity to compare directly our observed H$\alpha$ knots
with those separated by \citet{Koda2015} on their Subaru telescope
H$\alpha$ image of KK242 with the seeing of 0.8~arcsec. However, their
small angular size on the 2d spectrum indicates that their observed extent
is mainly due to the seeing.

Deep images of KK242 obtained with the {\it HST}/ACS in 2019 in the framework of
the SNAP program 'Every Known Nearby Galaxy' (Prop. 15922, PI R.B.
Tully) are available in the HLA\footnote{https://archive.stsci.edu}.
Stellar photometry of the resolved stars at the KK242 ACS images is
available at the Extragalactic Distance Database \citep[EDD;][]{Anand21}. They provide
us with the suitable data to get a deeper insight in the star-forming complex
covered by the BTA long slit.

We used both F606W and F814W  {\it HST}/ACS images of this program to identify
objects related to the H$\alpha$ emission in our spectrum and to estimate
their available physical parameters such as luminosity, size, colour.
Close in the position to the S knot, in both {\it HST} images there is a well resolved
almost round nebulosity with the diameter of $\sim$ 0.5~arcsec (or $\sim$15~pc).
No any reliable
star-like counterpart is visible within the nebulosity extent. On the other hand,
a more careful analysis (see Section~\ref{ssec:HST-objects}) of this {\it HST} image
reveals a blue star at $\sim$0.7~arcsec to the North (No.~5 in Fig.~1, right),
which could ionize the surrounding gas and, thus, to also contribute to the
H$\alpha$  emission in this region.

For the Northern  H$\alpha$ knot, the situation also appears to be complicated.
We discuss this issue in Section~\ref{ssec:HST-objects}.
The resulting 1d spectra are shown in the middle and bottom panels of Fig.~2.

\begin{figure}
\centering{
\includegraphics[width=6.0cm,angle=-0,clip=]{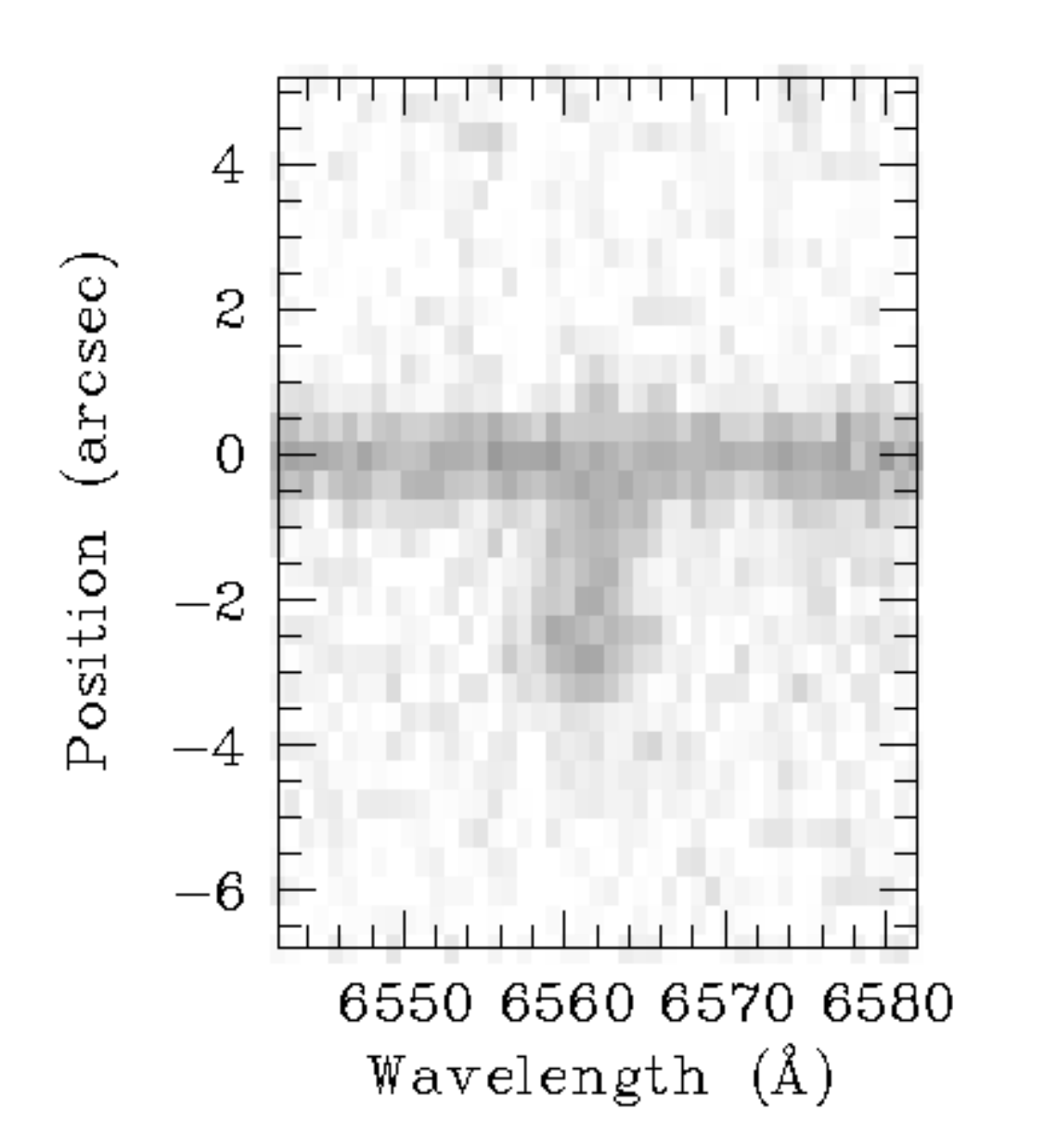}
\includegraphics[width=6.5cm,angle=-90,clip=]{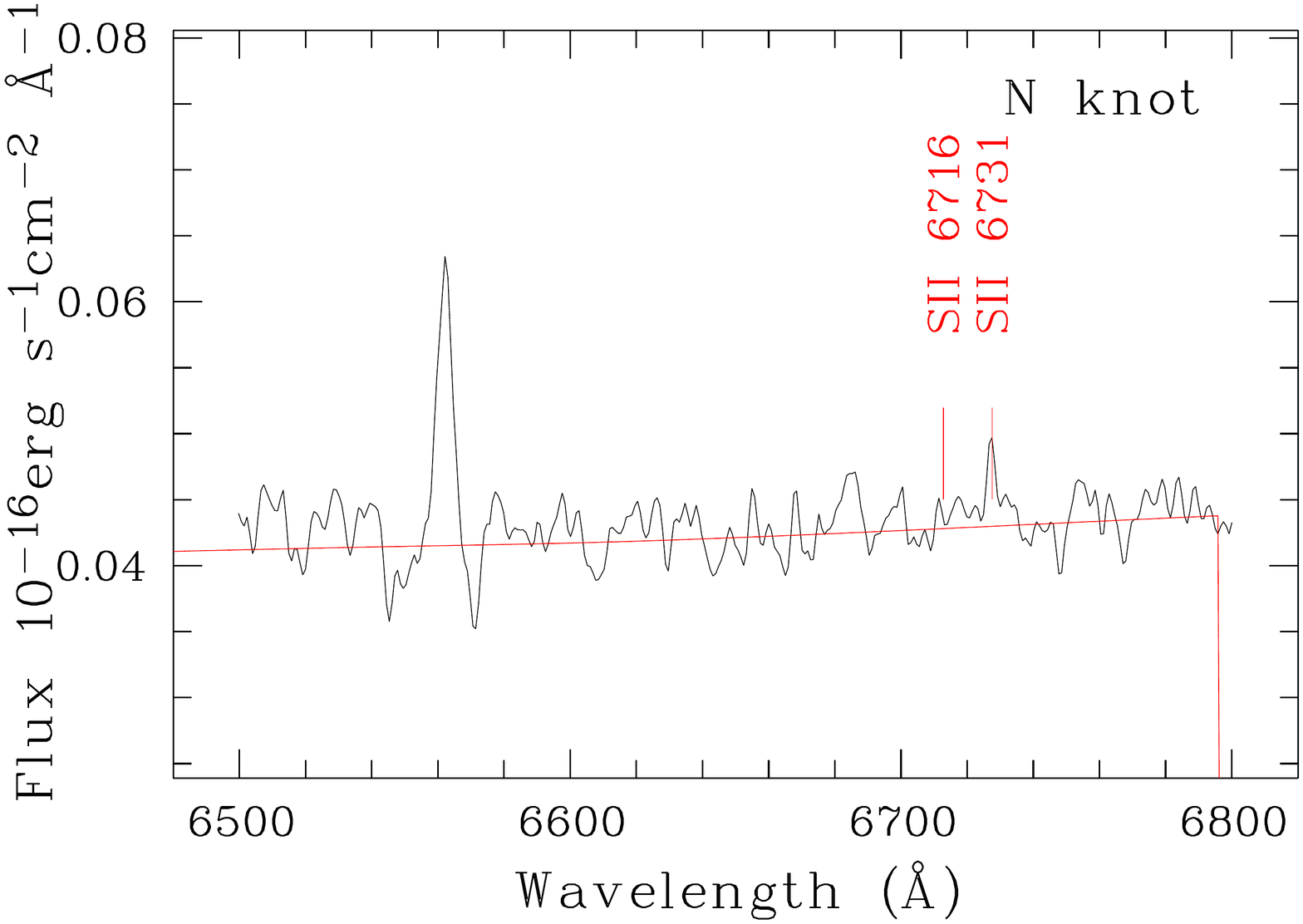}
\includegraphics[width=6.5cm,angle=-90,clip=]{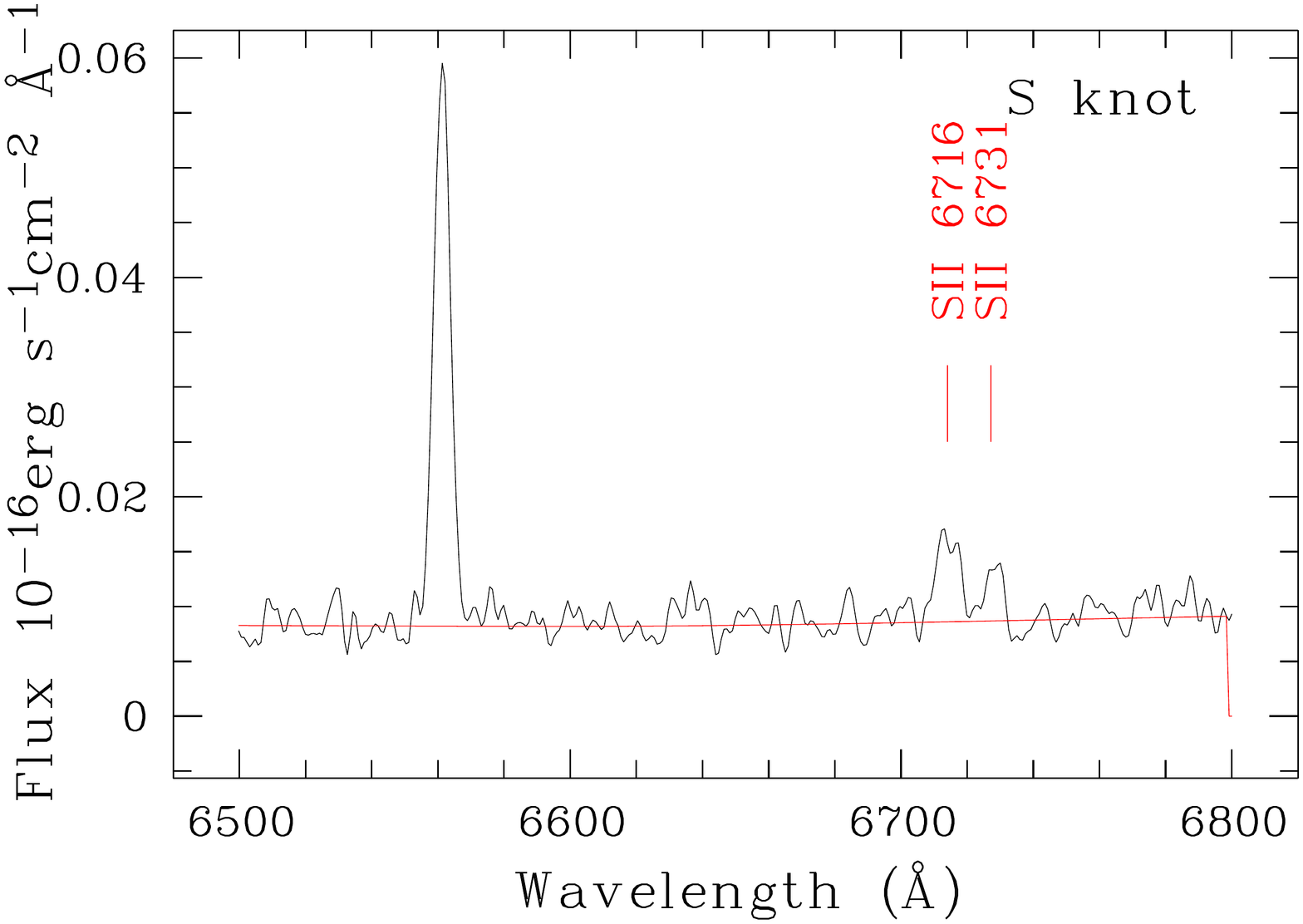}
\caption{BTA spectra for two H$\alpha$ knots in KK242.
{\it Top panel:} extract of the 2D spectrum in the range
6540--6580~\AA. North is up.
The distance along the slit between the peaks of two H$\alpha$ knots is about
$\sim$ 2~arcsec. Peak of H$\alpha$
emission for the N knot is shifted to the S by $\sim$0.7~arcsec relative to
the position of the continuum peak. See  Sec.~\ref{ssec:HST}.
{\it Middle  panel:} Average of the three 1d spectra
of the N knot with the 'strong' continuum and H$\alpha$
 flux twice fainter than for the S knot. For the 'undetected' [S{\sc ii}] doublet,
we show also the best two-gaussian fit in Fig.~3 (left),
which is used to estimate the  flux in this doublet.
See details in the text.
{\it Bottom panel:} The similar average 1D spectrum of the S knot
with the weak continuum, stronger H$\alpha$ and [S{\sc ii}].
 } }
\label{fig:Spectra}
\end{figure}

\begin{figure}
\centering{
\includegraphics[width=4.0cm,angle=-0,clip=]{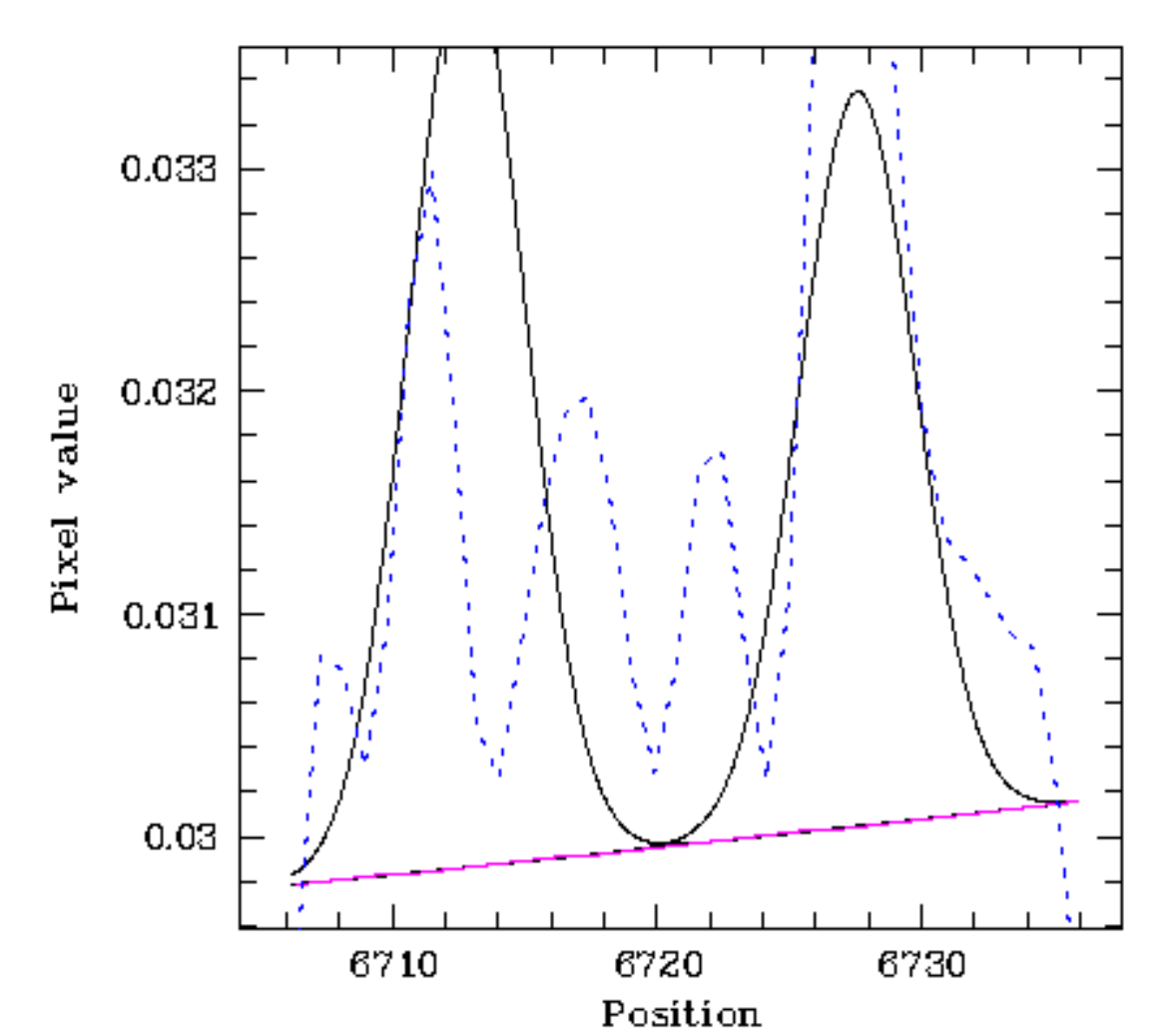}
\includegraphics[width=4.0cm,angle=-0,clip=]{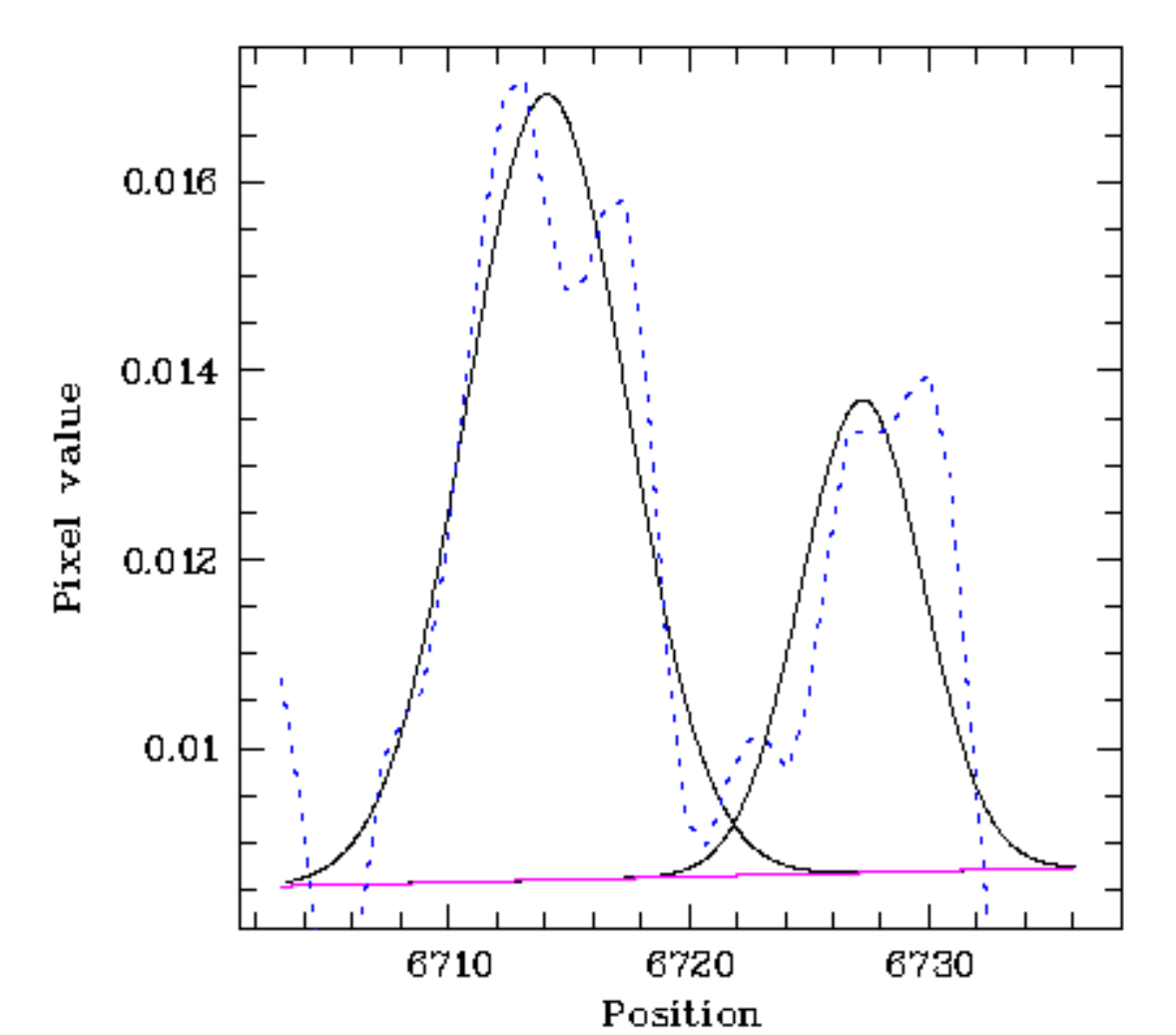}
\caption{Two-gauss fitting of the region of [S{\sc ii}]$\lambda\lambda$6716,6731
doublet in KK242. Dotted lines show the observed signal. Solid lines show
the fitting.
See comments on the gauss fitting in the text.
{\it Left-hand panel:} The N knot.
{\it Right-hand panel:} The S knot.
 } }
\label{fig:SII-fit}
\end{figure}

\begin{figure}
\centering{
\includegraphics[width=6.5cm,angle=-90,clip=]{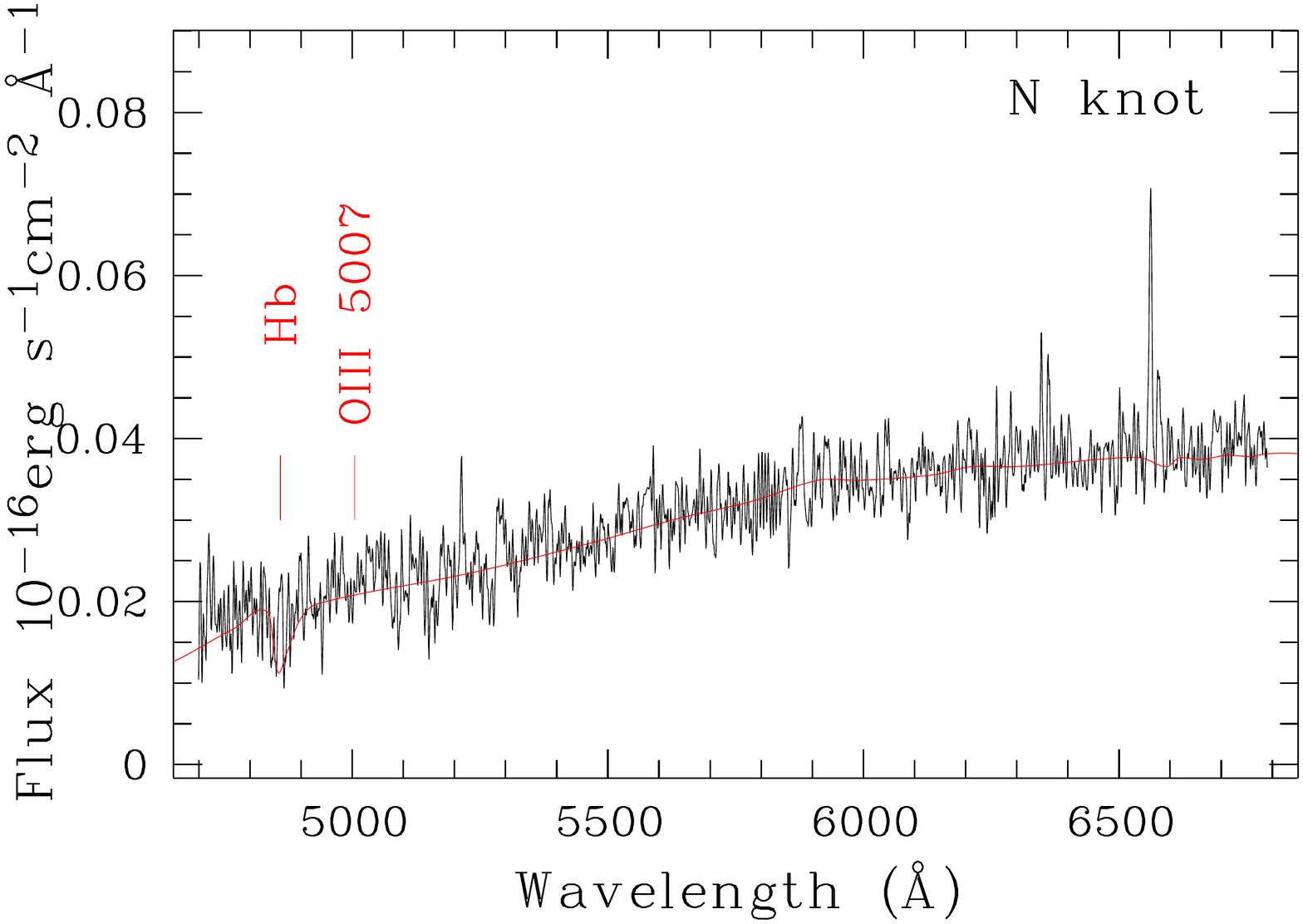}
\includegraphics[width=6.5cm,angle=-90,clip=]{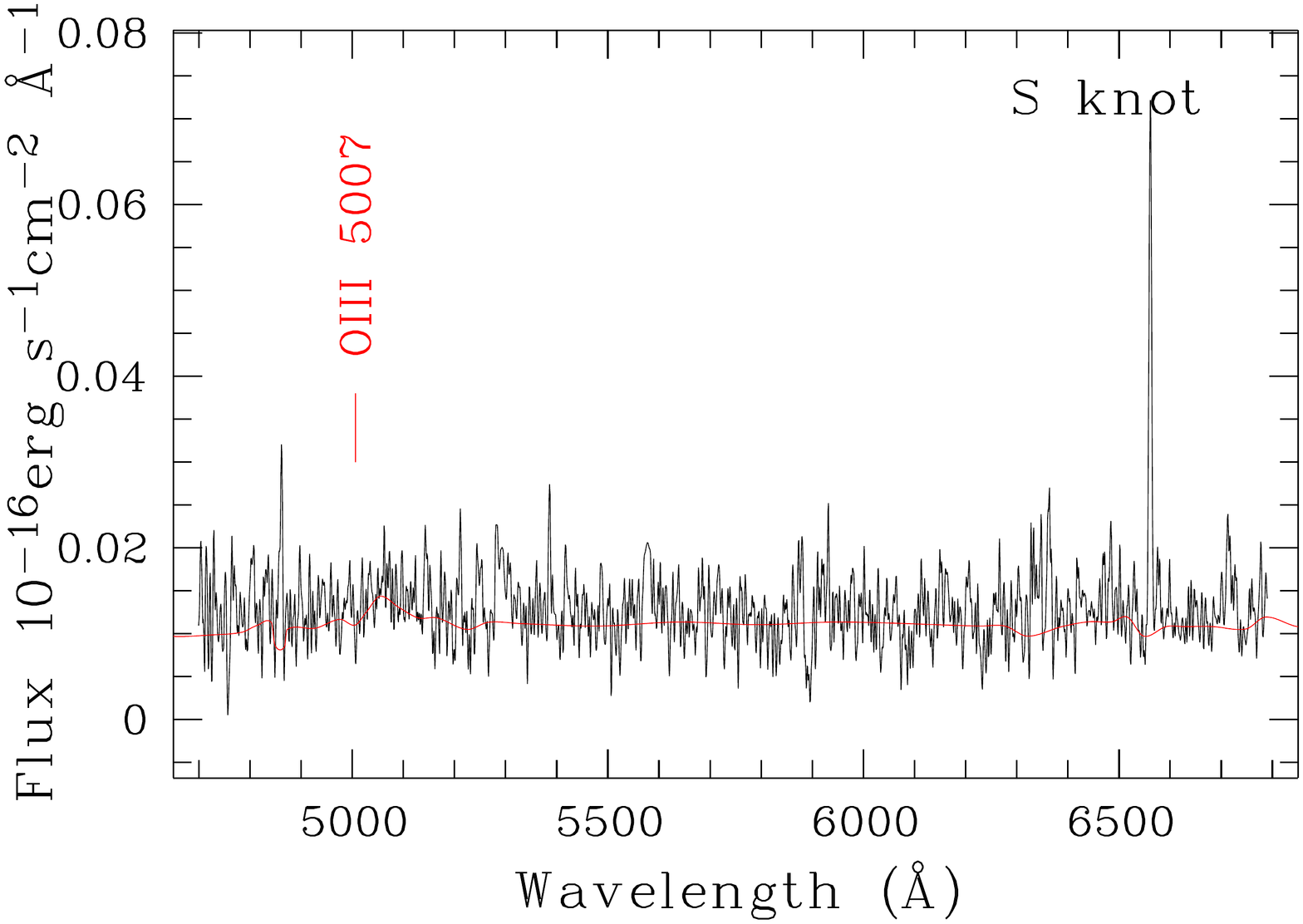}
\caption{BTA spectra of 2021.11.05 of KK242 for the N and S knots in the
range from H$\beta$ to H$\alpha$  and [S{\sc ii}]$\lambda\lambda$6716,6731.
{\it Top panel:} The N knot including part of continuum of star No.~1
(the extracted range along the slit of $\Delta$X = [--0.3,--1.5 arcsec]
in Fig.~6. {\it Bottom  panel:} The S region including continuum of the nebula
(the extracted range of $\Delta$X = [--1.5,--3.5 arcsec] in Fig.~6.
 } }
\label{fig:sco2-spectra}
\end{figure}

\begin{table}
\centering{
\caption{Average (top) and separate nights (bottom) parameters of S and N knots in KK242.
Line fluxes in 10$^{-17}$ erg~cm$^{-2}$~s$^{-1}$; FWHM in \AA. }
\label{t:Intens_line}
\begin{tabular}{l|c|c|c|c} \hline
\rule{0pt}{10pt}
				    & \MC{2}{c|}{S knot (mean)}   & \MC{2}{c|}{N knot (mean)}  \\ \hline
$\lambda_{0}$(\AA) Ion              & F($\lambda$)  & FWHM$\dagger$& F($\lambda$) & FWHM$\dagger$  \\ \hline
 6563\ H$\alpha$\                   & 2.98$\pm$0.05 &5.8$\pm$0.1  & 1.63$\pm$0.07 &5.1$\pm$0.1\\
 6717\ [S\ {\sc ii}]\               & 0.44$\pm$0.11 &7.9$\pm$0.7  & 0.11$\pm$0.15 &5.1~~~~~~  \\
 6731\ [S\ {\sc ii}]\               & 0.23$\pm$0.09 &6.0$\pm$1.0  & 0.07$\pm$0.13 &5.1~~~~~~  \\  
EW(H$\alpha$)\ \AA\                 & \MC {2}{c|}{33.3$\pm$0.7}   &  \MC {2}{c|}{5.5$\pm$0.3} \\ \\
 2021.11.05  & \MC{2}{c|}{S knot} & \MC{2}{c|}{N knot}   \\  
$\lambda_{0}$(\AA) Ion              & F($\lambda$)  & FWHM        & F($\lambda$)  & FWHM       \\ \\  
 4861\ H$\beta$\                    & 1.47$\pm$0.22 &6.5$\pm$0.9  & 0.44$\pm$0.11 & 8.3$\pm$1.9 \\
 5007\ [O\ {\sc iii}]\              & $<$0.30 (2$\sigma)$ & ...   & $<$0.20 (2$\sigma)$ & ...         \\
 6563\ H$\alpha$\                   & 3.62$\pm$0.15 &5.9$\pm$0.1  & 1.74$\pm$0.11 &6.2$\pm$0.2 \\ \\  
Rad. vel.\                          & \MC {2}{c|}{--65$\pm$10 \kms}& \MC {2}{c|}{--44$\pm$14 \kms} \\  \\ 
 2020.11.11  & \MC{2}{c|}{S knot} & \MC{2}{c|}{N knot}   \\  \\  
Rad. vel.\                          & \MC {2}{c|}{--80$\pm$10 \kms}& \MC {2}{c|}{--59$\pm$14 \kms}\\  \\ 
2022.07.29                       & \MC{2}{c|}{S knot} & \MC{2}{c|}{N knot} \\  \\ 
Rad. vel.\                          & \MC {2}{c|}{--87$\pm$10 \kms}& \MC {2}{c|}{--61$\pm$14 \kms}\\ \hline 
\multicolumn{5}{p{8.0cm}}{
{\it Notes.}  $^{a}$ For S knot, the average FWHMs for all three lines are taken on two observations with SCORPIO-1 as having
the better spectral resolution.
For N knot, the similar average only for the H$\alpha$ line is adopted. For the low-signal [S\ {\sc ii}] lines,
the two-gauss fitting was performed with the FWHM equal to that of H$\alpha$.
}
\end{tabular}
}
\end{table}

\begin{figure}
\centering{
\includegraphics[width=6.5cm,angle=-90,clip=]{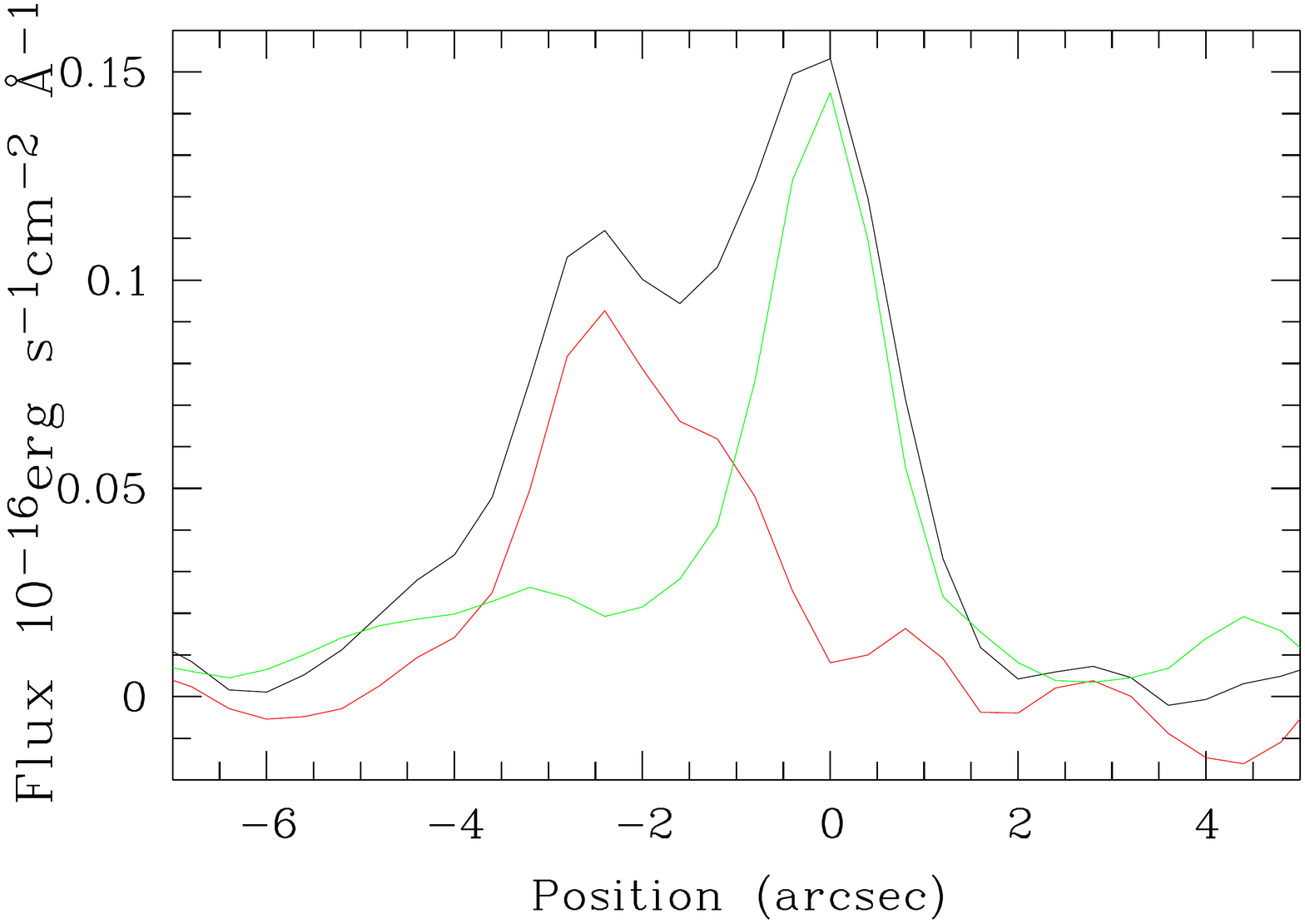}
\includegraphics[width=6.5cm,angle=-90,clip=]{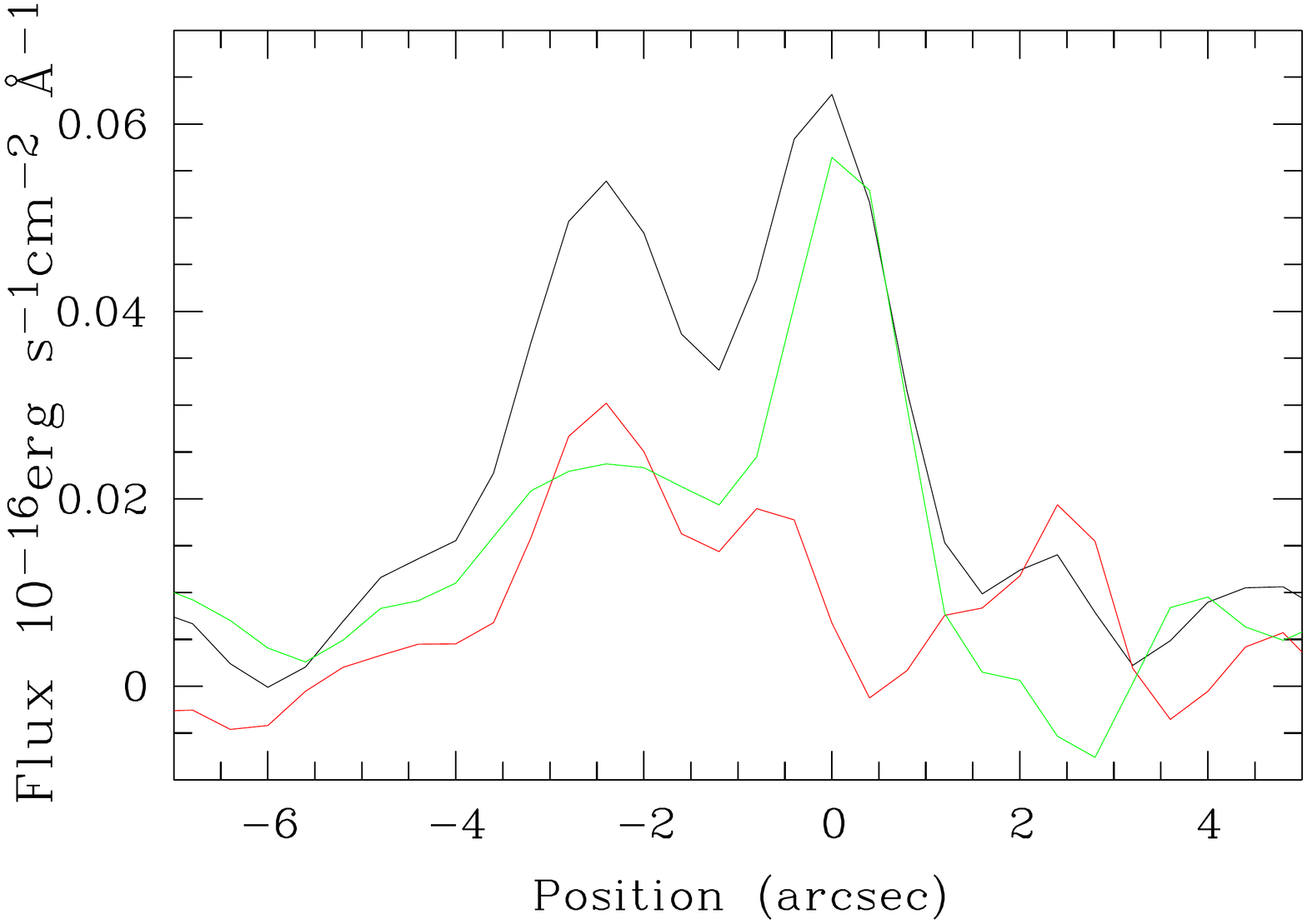}
\caption{H$\alpha$ line and the adjacent continuum emission distribution along
the slit in KK242 spectrum on 2021.11.05. The position of the brightest continuum
peak (X = 0) corresponds to position of Star No.1 at the {\it HST} image in Fig.~1
 (right panel).
{\it Top panel:} Net H$\alpha$ (red), sum of the adjacent continuum and H$\alpha$
(black), the adjacent continuum (green). The secondary 'peak' of continuum is at $\sim$3
arcsec to the South, is close to the positions of the Nebula and star No.~5 and looks
to be extended further to the South.
{\it Bottom  panel:} Same for the region of H$\beta$ emission.
 } }
\label{fig:along-slit}
\end{figure}

\section{Results}
\label{sec:results}

\subsection{Emission line parameters of KK242}
\label{ssec:emis.lines}

The results of emission line measurements and analysis for the
average spectra of all three nights are presented
in Table~\ref{t:Intens_line}. For the S and N emission-line 'knots',
we present the absolute fluxes (in units of 10$^{-17}$ erg~cm$^{-2}$~s$^{-1}$)
for H$\alpha$ and [S{\sc ii}] doublet and their widths (FWHM), as well as
the equivalent widths of H$\alpha$ and the measured heliocentric radial
velocities with their uncertainties.

We notice that the measured line widths in the spectra of the S knot are
somewhat broaden relative to the instrumental FWHM $\sim$ 5.1~\AA, as measured for
the H$\alpha$ line in the N knot.
While for the lines of the [S{\sc ii}] doublet, the S-to-N is lower,
there is also a hint on their broadening in the S knot.
The spectral resolution for the
spectrum on 2021 November 5, is worse ($\sim$6.2~\AA\ as indicated by the
the FWHM of H$\alpha$ in the N knot). Also, for the S knot, this parameter is
of the lower accuracy.
Hence, we use for the estimate of the broadening of
H$\alpha$ in the S knot only the data from the first and third observations conducted
with SCORPIO-1 with the same set-up.
This broadening from 5.1 to 5.8$\pm$0.1~\AA\ implies that the intrinsic value
of FWHM $\sim$ 2.8$\pm$0.2~\AA, or
the respective velocity width of $\sim$ 128$\pm$9~\kms.

The radial velocities measured on H$\alpha$ for the S and N knots,
averaged on the three independent datasets, are respectively, --77 $\pm$ 10~\kms\
and --55 $\pm$ 10~\kms.
The radial velocities for the both knots seem to show the systematic difference of $\sim$20~\kms\
in all three spectra. We have no at the moment  additional arguments to treat
one of them to be a more representative of the galaxy radial velocity.
Therefore,  we adopt their average of --66 $\pm$ 10~\kms\
as the robust estimate of KK242 radial velocity in the line H$\alpha$.

The line [N{\sc ii}]$\lambda$6584 is not detected. The upper limit on
its flux F($\lambda$) in the S knot, adopted as 2$\sigma$(noise) in
the adjacent continuum multiplied by the instrumental FWHM = 5.0~\AA,
is 0.2 $\times$ 10$^{-17}$
erg~cm$^{-2}$~s$^{-1}$. This is a factor of three smaller than the flux
of the [S{\sc ii}] doublet.

The relative flux of [S{\sc ii}] doublet to that of H$\alpha$ is a sensitive
parameter for checking the possible range of gas metallicity in KK242.
For the S knot, the contribution of the emission from the nebula, a probable
SN remnant, can substantially exceed the expected contribution from an \HII\ region
excited by a nearby hot massive star (see discussion in Sect.~\ref{ssec:HST}).
For the N region, we have an unambiguous case, in which only the nearby hot stars provide
the ionizing radiation, so that [S{\sc ii}] lines should reflect the abundance
and physical conditions characteristic of normal \HII\ regions.

The comparison of the three individual spectra of the N knot indicates that
the strength of the [S{\sc ii}] doublet relative to that of H$\alpha$ varies
by a factor of 2--3. This means that these faint lines are in fact at the level of
about one $\sigma$ of the noise or so. Therefore, discussion of [S{\sc ii}] doublet
in the individual spectra is irrelevant.
To derive a more reliable estimates of the strength of the [S{\sc ii}] doublet,
we obtained the weighted mean of all three individual spectra with the weights
adopted as 1/$\sigma_{\rm noise}^{2}$. Here the noise was estimated on the wavelength range
between H$\alpha$ and [S{\sc ii}] doublet. As one can see in Figure~2 (middle) and
Figure~3 (left), the signal in the individual lines of the doublet in the N knot is
comparable to the noise spikes.

We employ the next approach to get the best approximation of the real line fluxes.
We use all \mbox{a priori} information on the line positions and widths  based on the measured
parameters of the H$\alpha$ line. We also adopt the low-density case and the related
flux ratio I(6716)I(6731) = 1.4. With these parameters fixed, we vary the amplitude of
the main peak to get the lowest value of the residuals.  The resulting line fluxes are
shown in the top part of Table~\ref{t:Intens_line}.
For further discussion, we use for the N knot the parameter of flux ratio
[S{\sc ii}]/H$\alpha$ = 0.18/1.63 = 0.11,
with the conditional uncertainty of 50\%, that is 0.11 $\pm$ 0.055.
For the S knot, the respective value of [S{\sc ii}]/H$\alpha$ is 0.67/2.98 = 0.225 $\pm$ 0.073.
We use this parameter in Section~\ref{ssec:O/H}.

In Fig.~4, we present a wider range spectra of both the N and S knots of KK242.
They show the absence of the [O{\sc iii}]$\lambda$5007 that is consistent with the very low
excitation in the discussed \HII\ regions.

For discussion of the possible interpretation of the visible nebular
emission, it is useful to examine the H$\alpha$ flux distribution along
the slit relative to the northern peak of continuum related to the brightest
red star at the {\it HST} image, star No.~1 (see the right-hand panel of Fig.~1).
In the top panel of Fig.~5, we show the flux distribution along the slit for
H$\alpha$-line emission (red), for the adjacent continuum (green) and for
their sum (black). To tie the relative positions of
the line emission to the position of the 'bright' red star No.~1 in
Fig.~1, we count X-axis coordinate in Fig.~5 from the position of the highest
peak of continuum which corresponds to this star.
Negative values of X-axis correspond to the direction approximately to the south.

One can see that H$\alpha$ emission has a two-peak distribution, both
shifted to the S relative to star No.~1. A two-Gaussian fit to the flux
distribution allows us to get their positions as X(S) = --2.5~arcsec
and X(N) = --0.7~arcsec.

Finally, the total flux of H$\alpha$ is adopted as a sum of the nominal
fluxes in S and N knots according to the data on 2021.11.05 as obtained
with the better seeing and a wider slit (bottom of Table~2), namely
F(H$\alpha$) = (5.36 $\pm$ 0.18) $\times$ 10$^{-17}$ erg~cm$^{-2}$~s$^{-1}$.
We expect that due to the loss of a fraction of H$\alpha$ emission falling
outside the slit, the real flux in this region can be a factor of 1.5--2 larger.
We address the comparison of this parameter with earlier data in
Section~\ref{ssec:HST}.

\subsection{{\it HST} objects in the SF region}
\label{ssec:HST-objects}

In Table~\ref{tab:HSTdata}, we present positions and parameters of the brightest
8 stars and of one nebula which fall within the central area of the BTA long slit.
They all can contribute to the visible continuum and the line emission.
All these objects are marked in Fig.~1 (right). This should help to identify
them in the course of the further discussion.
Their {\it HST}/ACS pixel coordinates and visible V and I magnitudes are taken from
the table KK242.phot.WEB at the EDD
site\footnote{https://edd.ifa.hawaii.edu/get$\_$cmd.php?pgc=4689184},
where parameters of all measured in KK242 636 stars are presented.

Their world coordinates are taken from the available {\it HST} fits images of this region.
For the nebula, we obtained our own aperture photometry with the radius of
0.25~arcsec, which encompasses the whole object's light. We linked its instrumental
magnitudes with those derived for star No.~1. From their difference, we derived the V and I
magnitudes of the nebula, presented in Table~\ref{tab:HSTdata}.  We further discuss objects
identified at the {\it HST} images in Sect.~\ref{ssec:HST}.

\begin{table*}
\caption{{\it HST} based parameters of objects in the studied SF region of KK242}
\begin{tabular}{lcccccccc}
\hline
Object         & x       & y  & Coordinates (J2000)       & Vmag   &  Imag  &  M$_{\rm V,0}$ & (V--I)$_{\rm 0}$ & Approx. \\
	       & px      & px & relative to 17:52, +70:08 &        &        &                &                  & classif.   \\
\hline
No.~1          & 2128.7  & 3087.6 & 47.084, 11.25   & 22.14$\pm$.01 & 20.60$\pm$.01 & --6.98 &  +1.51$\pm$.01 & RHeB \\
No.~2          & 2120.5  & 3085.6 & 47.101, 10.87   & 24.82$\pm$.03 & 24.82$\pm$.05 & --4.30 & --0.02$\pm$.06 & BHeB \\ 
No.~3          & 2107.4  & 3082.2 & 47.128, 10.23   & 25.79$\pm$.05 & 26.15$\pm$.12 & --3.23 & --0.37$\pm$.13 & B0V  \\
No.~4          & 2095.9  & 3082.9 & 47.177, 09.70   & 26.23$\pm$.07 & 26.52$\pm$.16 & --2.79 & --0.32$\pm$.17 & B0V-B1V  \\
No.~5          & 2075.0  & 3077.9 & 47.211, 08.67   & 25.69$\pm$.05 & 25.88$\pm$.10 & --3.33 & --0.21$\pm$.11 & B0V-B1V  \\
No.~6          & 2099.8  & 3096.4 & 47.285, 10.15   & 25.63$\pm$.05 & 26.10$\pm$.12 & --3.39 & --0.49$\pm$.14 & O9V  \\ 
No.~7          & 2092.7  & 3091.9 & 47.271, 09.74   & 24.71$\pm$.03 & 24.62$\pm$.04 & --4.41 &  +0.07$\pm$.05 & BHeB  \\
No.~8          & 2070.2  & 3087.0 & 47.315, 08.60   & 26.48$\pm$.07 & 25.09$\pm$.06 & --2.54 &  +1.37$\pm$.08 & RGB?  \\
Nebula         & 2062:   & 3073:  & 47.231, 07.94   & 25.12$\pm$.07 & 23.44$\pm$.03 & --4.00 &  +1.66$\pm$.08 & SNR  \\
\hline
 \multicolumn{9}{p{15.6cm}}{{\it Notes.}
1. For Nebula: coordinates of the centre and our aperture photometry with the radius of 5~pixels (0.25~arcsec).}\\
\multicolumn{9}{p{15.6cm}}{2. Only stars are shown falling within the region of H$\alpha$ emission with V $<$ 27.0~mag (M$_{\rm V,0} \lesssim$ --2.0). } \\
\multicolumn{9}{p{15.6cm}}{3. E(V--I) is adopted 0.02~mag.} \\
\end{tabular}
\label{tab:HSTdata}
\end{table*}

\section{Discussion}
\label{sec:discuss}

\begin{table}
\caption{Properties of KK242}
\begin{tabular}{lcc}
\hline
Property                  & Value        & Refs          \\
\hline
RA (J2000)                & 17 52 48.4   &   4            \\
Dec (J2000)               & +70 08 14    &   4            \\
Rad. velocity (H$\alpha$), \kms\ & --66$\pm$10  & 1             \\
Rad. velocity (\HI), \kms\ & --80$\pm$10  & 2             \\
$D_{\rm adopt}$           & 6.36$\pm$.20 & 3             \\
Distance mod. (mag)       & 29.02        & 2             \\
B$_{\rm tot}^1$ (mag)     & 18.64        & 4             \\
A$_{\rm B}$ (mag)         & 0.14         & 5             \\
$M_{\rm B}$(mag)          & --10.5       & 2             \\
$\mu_{\rm 0,B}$ (mag~arcsec$^2$)& 25.5   & 4             \\
S(\HI) (Jy~\kms)          &$\lesssim$0.1 & 2             \\
M(\HI) (in 10$^6$ M\sunn) & 1$\pm$0.5   & 2             \\
M(*) (in 10$^6$ M\sunn)   & 6.0          & 4            \\
\hline
\multicolumn{3}{p{8.0cm}}{{\it Notes.}
1. This work (weighted average of both knots). 2. \citet{KK242.HST}.
3. Adopted as a mean of distances to KK242 and NGC6503 from \citet{KK242.HST}.
4. \citet{Koda2015}, M(*) scaled to the distance of 6.36~Mpc.
5. The Milky Way extinction after \citet{Schlafly11}.
}
\end{tabular}
\label{tab:parameters}
\end{table}

\subsection{Morphology of the SF region and related issues}
\label{ssec:HST}

The long-slit position for spectroscopy of the \HII\ region in KK242 was
determined based on the published H$\alpha$ images available at that time
\citep{Koda2015}.
The slit PA = --15\degr\ was also chosen to put the slit closer to the visible
elongation of the \HII\ region in order to increase the chances to catch
a knot with a higher excitation.
As the analysis of the {\it HST} image in  Fig.~1 (right)
reveals, the slit properly covers all hot stars except the bluest one, no.~6.
For this object, accounting for the seeing of 1.1--1.3~arcsec, we probably lose
up to a half of the star and the related \HII\ region flux relative to
the other discussed stars.

Besides, as described at the end of Sect.~\ref{ssec:emis.lines},
the nominal measured flux of H$\alpha$ in this region is
$\sim (0.54 \pm 0.02) \times 10^{-16}$ erg~cm$^{-2}$~s$^{-1}$, with the possible
upwards correction by a factor of 1.5--2 due to the loss on the slit.
There are two independent estimates of this parameter (obtained from H$\alpha$ imaging,
in units of 10$^{-16}$ erg~cm$^{-2}$~s$^{-1}$) in papers by \citet{Koda2015}
(2.3 with the uncertainty of factor of two) and \citet{KK242.HST} (4.7$\pm$2.5).
Taking into account their large uncertainties and our upward correction of the
nominal flux of H$\alpha$, we suggest that the real flux of H$\alpha$ in this region
 is of $\sim$10$^{-16}$ erg~cm$^{-2}$~s$^{-1}$.

Our task is to connect the visible H$\alpha$ emission on the BTA long-slit
spectra with the exciting hot stars visible in this region at the {\it HST} image.
This allows us to better understand probable parameters of the SF episode
in this region.
Besides, this information will be helpful in the attempt of modelling
the observed nebular emission (e.g. with {\sc Cloudy} package, see the Appendix).

As the derived (V--I)$_{\mathrm 0}$ colours of stars in Table~\ref{tab:HSTdata}
indicate, the two brightest stars, no.1 and no.2 are not hot and therefore have no relation
to the observed nebular emission. No.1 is likely a Red Helium Burning (RHeB),
while no.2 and 7 with the colour (V--I)$_{\mathrm 0}$ $\sim$ 0.0 mag
are likely Blue Helium Burning (BHeB) stars \citep{McQuinn11, McQuinn12}.
Both types are products of the late evolution with the He-core burning of stars
with the intermediate masses of 2--15 M\sunn.
The measured absolute magnitudes of M$_{\rm V}$ = --7.0 mag for star no.~1 and
--4.3 and --4.4 mag for no.2 and 7,
evidence to the extended episode of SF in this region, with the duration of at
least of $\sim$100 Myr \citep{McQuinn11}.

Stars no.3, 4, 5 and 6 all are rather blue and luminous, seemingly
representing the main-sequence late O-type and/or early B-type stars. We use
the sample of O and B stars from the Wing of the Small Magellanic Cloud (SMC)
of \citet{Ramachandran19} for which these authors
present the observed and model physical parameters. Since this large sample of
OB stars have the nearest metallicity and thus are the best proxies to
KK242 massive stars, we use their M$_{\rm V}$ as a comparison for our
blue luminous stars. As one can see, the SMC OB stars show the substantial
scatter in their M$_{\rm V}$. Accounting for this information, the blue stars
found in the KK242 SF complex can be assigned to the range of O9V
to B1V. The bluest (hottest) star No.~6 is probably an O9V, while the
remaining, a bit redder stars, are probably of B0V--B1V type.

The positions of blue stars no.~3, 4 and 6 are close to the position
of the brightness peak of the Northern H$\alpha$ knot. Respectively, the position of
the brightness peak of the Southern H$\alpha$ emission is close to the position
of the blue star no.~5. On the other hand, it is also rather
close to the position of the adjacent red nebula, so that it can contribute
to the observed nebular emission of the S region as well.

Since the projected distances between the hot stars no.~3, 4 and 6
are of $\sim$20--30 pc, for the gas density N $\gtrsim$ 10~cm$^{-3}$, their
Str\"omgren radii should be smaller than their mutual distances, so it is
unlikely that they form a common \HII\ region.
However, the separate \HII\ regions around these stars  can contribute
to the total observed nebular emission in the northern H$\alpha$ knot due to the
insufficient angular resolution along the slit.
A similar situation takes place for the southern  H$\alpha$ emission. Here,
the hot star no.~5 is a clear candidate for the line emission from the related
\HII\ region.
However, the adjacent nebula, a probable SN remnant, can make a major
contribution.

Indeed, as one can see from Table~\ref{t:Intens_line}, the H$\alpha$ line
flux is a factor of two larger in the southern knot relative to that
in the northern knot. Taking into account that in the N region, we have three
hot massive stars (no.~3, 4, 6) with parameters close to those of star~no.5 in
the S knot, it is reasonable to assume that the H$\alpha$ luminosity of \HII\ region
excited by star~no.5 is lower (about factor of 2--3) than the respective
luminosity for the three \HII\  regions comprising the N knot.  From this consideration
it follows that the main contribution to the line and continuum emission of the S knot
comes from the nebula, a candidate SN remnant.

This coarse analysis shows that the observed spectra of this SF complex,
with the limited angular resolution typical of the ground-based observations,
represent rather complicated combination. So that their interpretation, even for
a higher S-to-N case, can be not  that straightforward. They can hope that
for the exceptional ground-based observational conditions, with a seeing of
$\lesssim$0.5~arcsec and with the appropriate slit width
oriented at a proper direction, one can disentangle the contribution of
various \HII\ regions and enable obtaining of the strong-line fluxes and
the subsequent estimates of the gas metallicity. More chances for the
detection of the strong oxygen lines are expected for the \HII\ region around
star no.~6, the hottest of all massive stars in KK242, judging on its V--I colour.
Also, the projected distance of $\sim$0.7~arcsec between star no.~5
and the centre of the nebula should be sufficient to disentangle emission
lines of the two objects.

\subsection{Environment}
\label{ssec:environs}

The dwarf galaxy KK242, as well as its host spiral galaxy NGC6503, resides
within a  nearby void, No.~22  (also named Dra-Cep) in the list of \citet{PTM19}.
Their distance to the nearest luminous galaxy NGC6946 is D$_{\rm NN}$ = 2.25~Mpc.
The void is a bit flattened spheroid with the large diameter of 21~Mpc. The
centre of the void is at the distance of 13.9 Mpc from the Local Group, in the
direction of RA = 20.4~h, Dec. = +71\degr. The Dra-Cep void is situated above
the supergalactic plane SGZ=0, being adjacent at SGX $\sim$ +10 to +20 Mpc to
the largest nearby void Oph-Sgr-Cap which includes the well known Local (Tully)
Void (see illustration in fig.~A5 of \citet{PTM19}). Fifty galaxies reside within
the void boundaries \citep{PTM19}. Of them, 44 are classified as the 'inner' void
galaxies, defined as those with the distance to their nearest luminous neighbour
$D_{\rm NN} \geq$ 2.0 Mpc. Both KK242 and NGC6503 are assigned to the 'inner'
void galaxies.

\subsection{Evolutionary scenarios and gas metallicity}
\label{ssec:O/H}

\subsubsection{KK242 as a void dwarf of transition type}

The phenomenon of dTr is not well understood. The assignment of a dwarf
to the transition type is a purely phenomenological.
While these dwarfs have in general the substantial amount of gas, comparable
with that in dIrrs, their current star formation as traced by the H$\alpha$
emission of the related \HII\ regions, is (almost) quenched.
Five the nearest and seemingly the best studied examples of dTrs (or dIrr/dSph)
are found within the Local Group and reviewed by \citet{Mateo1998}.
\citet{Skillman2003} added to them three similar dwarfs in the Sculptor group and
via the analysis of the {\it HST} CMDs addressed the nature of "transition" dwarfs.
They conclude that the examined  dTrs are similar on the gas content to dIrr and
are 'found preferentially among the lowest luminosities and nearer to spiral galaxies.
Their appearance thus is caused by the temporary interrupted star formation.
However, the tidal effects of massive hosts also may play a role'.

Later \citet{Weisz2011}, with deep {\it HST} data, studied SF histories (SFHs)
of 60 dwarfs within 4 Mpc. Of them, 12 are classified as dTrs. One of their
conclusions is that despite the large diversity, the mean SFHs of dIrrs, dTrs
and dSphs are similar over most of cosmic time, with the clearest difference
between the three only during the most recent 1 Gyr.
 They also conclude: 'In terms of their environment, SFHs and gas fractions,
the majority of the dTrs appear to be low-mass dIs that simply lack H$\alpha$ emission,
similar to the LG dTr DDO210. However, a handful of dTrs have remarkably low
(but detectable) gas fractions, suggesting that they nearly exhausted their gas supply,
analogous to the LG dTrs such as Phoenix and LGS3.'

As mentioned in Section Introduction, to date, the great majority
of the known dTrs are found in groups and near massive hosts \citep{KK258}.
This can be, at least partly, due to the observational selection effects,
since for the faint isolated dwarfs of this type, the determination of their
radial velocity, similar to the case of dEs, is a difficult task.

One can think on the different origin and evolutionary scenarios of dTr objects,
that can be related, in particular, to their global environment.
For the majority of the currently known dTrs, the simplest scenario assumes their
relation to 'normal' dIrrs with the lowest baryonic mass, in which the intermittent
SF occurs with the duty cycle larger than (a few) tens Myr.
For the alternative scenarios, in particular, for gas-poor dTrs, there are various
options. For example, a normal dIrr progenitor could lose the major part of its gas
due to the close passage near a more massive host \citep[e.g.,][]{DiCintio2021}.
Another option is a long-ago formed dSph/dE which was recently rejuvenated due to the
gas accretion. As far as we aware, the variant of the metal-enriched gas accretion
from the outer parts of a massive host to a dwarf companion is not yet modelled.
That is one can not describe, at which circumstances this will occur, if at all.
On the other hand, in voids, where gas velocities in filaments are low \citep{Aragon2013},
the unprocessed gas accretion from filaments to small dwarfs can probably work.

Coming back to the properties of KK242, we notice, that on the low gas content it
resembles the minority of dTrs, which 'nearly exhausted their gas supply', in difference
to the main group of dTrs.
Due to the similar SFHs of dIs and dTrs, with except of the last 1 Gyr \citep{Weisz2011},
one of the evolutionary scenarios for dTr objects involves the late gas removal and
the related star formation quenching. Therefore, one can expect that the progenitors
of gas-poor dTr objects have been evolving most of their lifetime similar to
dIrrs with the same mass. Then, if the metallicity of the removed gas was
typical of the gas in the whole galaxy, the remaining gas should be
representative of the previous secular evolution. In this scenario, it is probable
that the  gas metallicity in dTrs is similar to that in dIs with the same stellar
mass and luminosity.

\subsubsection{Gas metallicity as expected from the global parameters}

The gas metallicity of late-type galaxies in the LV follows the trend
described by the relation of 'O/H versus M$_{\rm B}$' from \citet{Berg12}.
The respective linear regression reads as 12+log(O/H) = 6.272 -- 0.107 $\times$ M$_{\rm B}$,
with the rms scatter of log(O/H) of 0.15~dex. It extends over the range of
M$_{\rm B}$ = [--9.0,--19.0]. The great majority of this LV reference sample
belongs to typical groups and their close environs. As shown in
\citet{PaperVII,XMP-BTA},
the late-type dwarfs in the nearby voids have, on average, the reduced
values of log(O/H) by 0.14~dex (or by $\sim$30 percent, with the
rms scatter of 0.18~dex) relative to this reference relation. This
finding was interpreted as an evidence of the slower galaxy  evolution
in voids. Consistently with this idea, void galaxies have also
the elevated \HI\ content, on average by 40 percent \citep{PaperVI}.

Since KK242 is not a typical late-type dwarf, the above statistical
relations between O/H and blue luminosity or stellar mass, derived
in \citet{Berg12}, may be not directly applicable to it.
Those relations are assumed to reflect the specifics of the secular evolution
of disc galaxies in the wide range of baryonic mass.

It is interesting to compare its gas metallicity with the other dwarfs of
the same blue luminosity. We first derive the expected gas O/H in KK242 if
its \HII\ region(s) metallicity obeys the above reference relation for the
Local Volume late-type galaxies from \citet{Berg12}.
For M$_{\rm B}$(KK242) = --10.5~mag, its expected 12+log(O/H) = 7.40 $\pm$ 0.15~dex.
\citet{Koda2015} present the estimate of the total stellar mass for KK242
for their adopted distance of 5.27~Mpc.
Accounting for the scaling due to the increased distance by a factor of 1.2,
we adopt it as log(M$_{*}$) = 6.78~dex.

We can use this log(M$_{*}$) for an alternative estimate
of the gas O/H, based on the similar relation from \citet{Berg12}, namely:
12+log(O/H) = 5.61 +0.29 $\times$ log(M$_{*}$),  with the rms scatter of
$\sigma$ = 0.15~dex. This gives the value 12+log(O/H) = 7.58$\pm$0.15~dex.
Taking the average of the two independent estimates (7.40 and 7.58), we adopt
the expected value of gas O/H for a typical dIrr with those M$_{\rm B}$ and
stellar mass, as 12+log(O/H) = 7.49$\pm$0.10~dex.

If we take into account that the secular evolution of KK242 took place within
a void, then, as mentioned in the beginning of this section, the expected
value of O/H is lower, on average by 0.14~dex, that is,
of 12+log(O/H) = 7.35 $\pm$ 0.18~dex.

\subsubsection{Use of [S{\sc ii}] doublet to constrain gas O/H}
\label{ssec:OH_gas}

With a lack of information on the strong oxygen lines, our spectral data,
on the first glimpse, cannot be used to derive more or less reliable
empirical estimate of O/H. The only means to probe gas metallicity in
KK242 and to check, whether this is consistent with the above estimate of
the expected O/H, is the relative strength of [S{\sc ii}] doublet. Its
statistical relation with the parameter 12+log(O/H) can provide us, in
principle, with the possible range of the gas metallicity of KK242 and
help in the comparison with the gas metallicity expected within a particular
scenario in the previous section.

In the following discussion of the N knot, we adopt the flux ratio of
the [S{\sc ii}] doublet and H$\alpha$ as a weighted mean of the three
independent measurements, as shown in the top of Table~\ref{t:Intens_line},
namely 0.110$\pm$0.055. We adopt conditionally, for illustration, the error
at the level of 50\%. From the formal estimates, it's probably twice larger.
In the further comparison of the strength of [S{\sc ii}] doublet to galaxies with
known O/H, we need its ratio to the flux of H$\beta$. We adopt it from the typical
flux ratio of H$\alpha$ and H$\beta$ for the Case~B recombination, of $\sim$2.8.
The latter is consistent with the flux ratio of the N component as visible in
the intensity cuts along the slit in Fig.~5 (top and bottom).
Then, the respective parameter, called S2 (a ratio of [S{\sc ii}] doublet flux to
that of H$\beta$),  is adopted for further to be equal to 0.31+-0.155.
That is the most probable range of S2 is [0.155,0.465].
The respective value of lg(S2)=--0.509, with the most probable range of [--0.810,--0.333].

The parameter log(S2) can be used in principle for  comparison with the
statistical data compiled by \citet{Counterpart2012} for various \HII\
regions in galaxies with the wide range of O/H.  In Fig.~6, we plot the relation
between the parameter log(S2) and 12+log(O/H) for a subsample of 161 data points from
the compilation by \citet{Counterpart2012} for all \HII\ regions with 12+log(O/H)
(the direct method) in the range of $\sim$ 7.1 -- 8.0~dex. The vertical solid black
and blue dashed lines mark the expected value of gas O/H for the absolute
blue magnitude and stellar mass of KK242 and its  \mbox{$\pm$1~rms}
corridor for the case when the gas metallicity (O/H) of KK242
obeys the reference relation for late-type galaxies from \citet{Berg12}.

As discussed above, for the S region, the H$\alpha$ emission appears to be the sum
of two components. The first is an \HII\ region excited by the star no.~5, a probable
star B0V-B1V, and the second is the emission from a round red nebula without a detectable
exciting central star. As we argued above, the main contribution to
the emission of this region is due to radiation from the nebula.
The profile of H$\alpha$ line in this region looks broaden by the amount
corresponding to the intrinsic FWHM $\sim$ 128~\kms. The latter corresponds to
a shell expansion with the characteristic velocity of $\sim$64~\kms. Therefore,
it is very likely that the round nebula is a supernova remnant (SNR) with an age
of less than 1~Myr.

It is well known that the ratio of [S{\sc ii}]$\lambda\lambda$6716,6731  flux
to that of H$\alpha$ (hereafter, [S{\sc ii}]/H$\alpha$) is enhanced
in the optical spectra of SNR due to the shock excitation.
The often used empirical criterion to assign the observed emission
to the shock-excited is [S{\sc ii}]/H$\alpha >$ 0.4. However,
\citet{Kopsacheili20} from the analysis of the shock-excitation models
based on the package MAPPINGS~III \citep{MAPPINGS} show that this parameter can be
substantially smaller for the subsolar gas metallicities and for the shock velocities
less than 200~\kms. Therefore, the observed in the Southern knot ratio
[S{\sc ii}]/H$\alpha \sim$ 0.225, accounting for a low gas metallicity and
a 'small' shock velocity, is consistent with the expected in the models.
Therefore, for the S region, we can not use this ratio for comparison with
the statistical data for normal \HII\ regions.

\begin{figure*}
\centering{
\includegraphics[width=12.0cm,angle=-90,clip=]{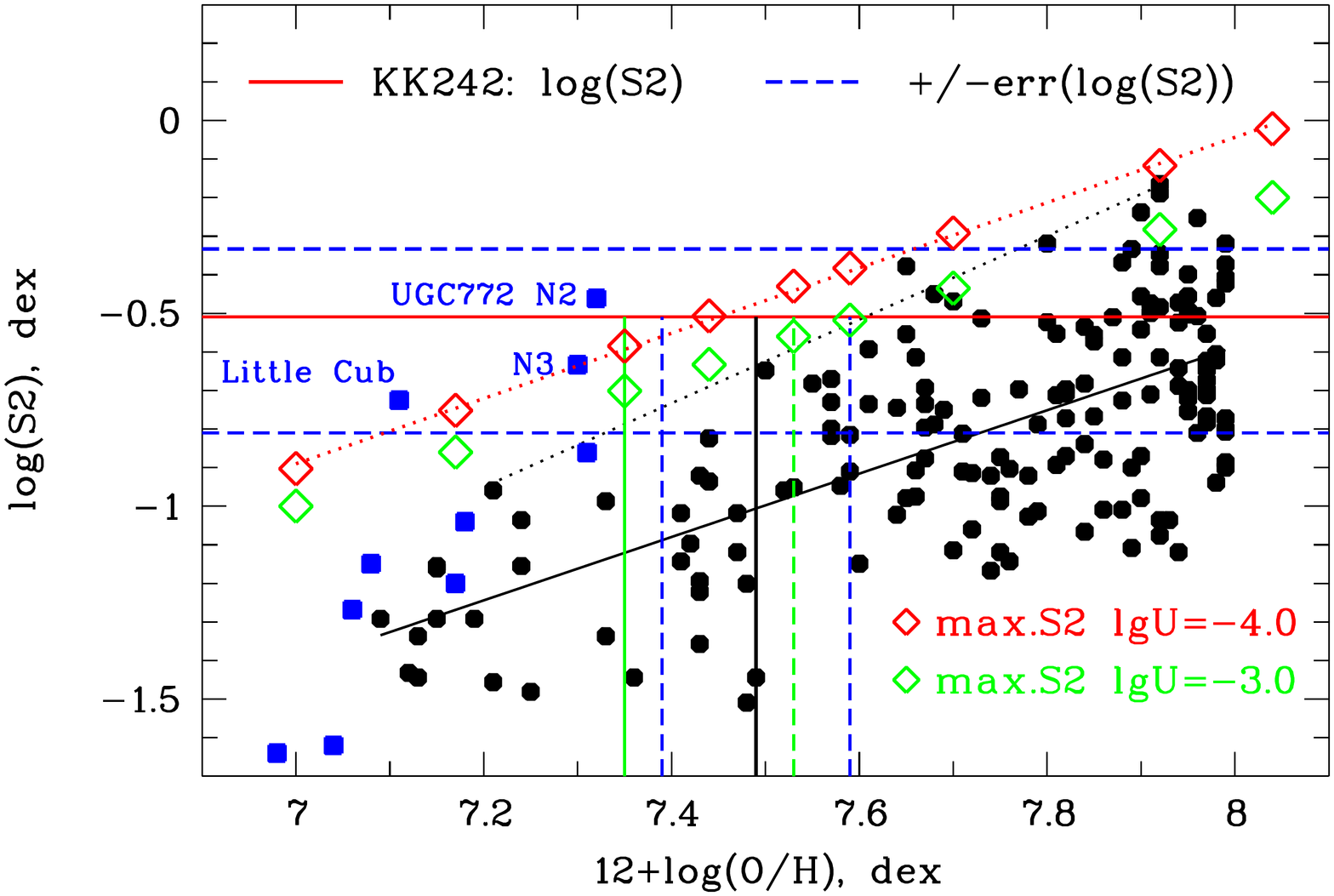}
\caption{Plot of parameter log(S2) versus 12+log(O/H) for observed and model
\HII\ regions. Black octagons show points for 161 \HII\ regions with 12+log(O/H)(T$_{\rm e}$)
$ < $ 8.0. They are drawn from the compilation of the literature data in
\citet{Counterpart2012}.
Vertical solid black and blue dashed lines show the expected value of 12+log(O/H) and its
probable variance (7.49$\pm$0.10~dex), corresponding to M$_{\rm B}$ and stellar mass of KK242,
in the case if its O/H follows the reference relation 'O/H versus M$_{\rm B}$'
for the LV sample of \citet{Berg12}. Green solid and dashed vertical
lines show the probable O/H for similar galaxies residing in voids (7.35~dex)
and its +1$\sigma$ value (7.53~dex) \citep{XMP-BTA}. The horizontal solid red and two blue
dashed lines show the nominal value of log(S2) for KK242 Northern knot
and its $\pm$1$\sigma$ range.
We also draw the linear regression of log(S2) on log(O/H) for the above sample of
161 points (solid black) and its upper envelope (dotted black).
Red and green diamonds correspond to the maximal values of S2 for a given value of
O/H, which occur for models with the values of the ionisation parameter
 of lg(U) = --4.0 and --3.0, respectively.
They occur for the lowest considered T$_{\rm eff}$
of the exciting stars, of 25--30~kK, as illustrated in Fig.~A1 of
Appendix A. The red dotted line shows the linear approximation of
of positions of the red diamonds.
We also add 9 points with the direct values of
12+log(O/H) $\lesssim$7.32 dex (blue squares), published after 2012
(see references in the text), to illustrate the large scatter of
the parameter S2 for the low values of O/H in
comparison to the limited data from \citet{Counterpart2012}. See text
for details and discussion of this figure.
 } }
\label{fig:lgS2}
\end{figure*}

For the N region, we adopt that the nebular emission, visible on the slit, is
the sum of emission  of three normal \HII\ regions excited by the hot stars
no.~3, 4 and 6.
In Fig.~6, the nominal value of log(S2)(KK242) = --0.509  for the N knot
is shown by the red horizontal line, while the lines, corresponding to
+1$\sigma$ (--0.333) and --1$\sigma$ (--0.81)   are shown by the dashed blue lines.

As well seen in Fig.~6, the nominal value of log(S2)(KK242), being directly compared to
the data points from \citet{Counterpart2012},
corresponds to rather wide range of 12+log(O/H) $\sim$ 7.8 $\pm$ 0.2~dex.
The latter is significantly larger than the O/H expected for its
very low M$_{\rm B}$ and M${*}$. This apparent inconsistency needs explanations
and a discussion.

Since the used data from \citet{Counterpart2012} is based on \HII\ regions with
the directly measured O/H, they all have the well detected line [O{\sc iii}]4363.
According to the {\sc Cloudy} models discussed in the
Appendix, this case corresponds to \HII\ regions with the sufficiently large value
of the ionization parameter, namely to log(U) $\gtrsim$ --3.0. According to the same
model grids, for a given gas metallicity, the parameter S2 can vary substantially for
the range of T$_{\rm eff}$ of the exciting stars of [25, 50] kK and the range
 of log(U) = [-4.0,-1.0].

To discuss more general cases, we draw in Fig.~6, in addition to the observed
\HII\ regions with the direct O/H from \citet{Counterpart2012},
the model-predicted values of S2 for a wider range of log(U) and T$_{\rm eff}$,
derived with the {\sc Cloudy} package in Appendix. Here, green and red
diamonds show the maximal values of log(S2) for nine values of 12+log(O/H) between
7.00 to 8.04 dex for log(U) = --3.0 and --4.0, respectively. These maximal values of
S2 correspond to the minimal values of T$_{\rm eff}$ from the grid with
T$_{\rm eff}$ = 25--50~kK with the step of 5~kK.

Coming back to the N region of KK242, we notice that its
nominal value of log(S2)=-0.51 corresponds to 12+log(O/H)=7.44~dex in the case
of this region is ionized by the lowest T$_{\rm eff}$ (25--30 kK) and the lowest
log(U) source ($\sim$ --4). This is well consistent with the expected void dwarfs metallicity
mentioned above of 12+log(O/H) = 7.35 $\pm$ 0.18 dex.

It is worth to noting that the sample of the reference \HII\ regions from
\citet{Counterpart2012} does not cover completely the real parametric space
in the plane 12+log(O/H), log(S2) for regions with the directly derived O/H,
at least for the lowest gas metallicities. We added in Fig.~6 ten data points
with the direct 12+log(O/H) $\lesssim$ 7.3~dex (blue squares), appeared in
the literature after 2011 in papers by \citet{Izotov12, Izotov18, Izotov19, Skillman13},
\citet{Hirschauer16, LittleCub, XMP-BTA}. Seven of them
fit well the region defined by the data from \citet{Counterpart2012} and its
extension to the lower O/H. However, the parameter log(S2) for
'Little Cub' and regions UGC772-N2 and N3
falls significantly higher than for the remaining majority.
In particular, the region UGC772-N2 has the S2 parameter close to that of the N knot  in
KK242 and very low 12+log(O/H) $\sim$ 7.3~dex.

\subsubsection{Probable parameters of exciting stars in KK242.
 Similar  case of Pegasus DIG}
\label{ssec:star_parameters}

To complete the discussion on the consistency of the observed parameter
S2 in the Northern H$\alpha$ knot in KK242 with the expected low gas metallicity,
of 12+log(O/H) $\lesssim$ 7.49 $\pm$ 0.10~dex, we examine the range of possible
parameters of the hot  massive
stars illuminating this H$\alpha$ region. We are interested whether their
effective temperatures and the ionising photon fluxes are consistent with
the {\sc Cloudy} package parameters resulting in the largest values of S2
for that low gas metallicity (red and green diamonds in Fig.~6).

The observed parameters of the related stars nos.~3, 4 and 6 are
summarized in Table~\ref{tab:HSTdata} and discussed in Section~\ref{ssec:HST}.
We use the results of modelling of a large sample of massive OB stars in SMC
presented in \citet{Ramachandran19}. The metallicity of this sample is
the closest to that expected for gas and young stars in KK242. Thanks to
the good statistics and the large set of modelled physical parameters,
this allows one to use the average parameters for a star of the given
spectral class as well as to understand the real range of their scatter.

The rough estimate of the expected parameter log(U) in the considered region
can be derived taking the typical flux of the ionising photons
of B0V stars in the SMC, presented in \citet{Ramachandran19}, as follows:
log(Q) = 47.26$\pm$0.3 and the effective temperature
of T$_{\rm eff}$ = 30$\pm$1~kK.\footnote{We notice that these parameters differ
substantially downwards from the adopted in Table~2.3 in the monograph
of \citet{OF2006}.}
The expected gas density is n$_{\rm e} \lesssim$ 10~cm$^{-3}$. This leads
to the Str\"omgren radius of $\sim$15~pc and the related value of lg(U) = --4.0.

As the examination of the {\sc Cloudy} grids in Appendix shows, for
12+log(O/H) = 7.35, the largest value of parameter S2 occurs at lg(U) = --4.0,
and for the lowest T$_{\rm eff}$ = 25--30 kK, reaching the value of 0.26.
Similarly, for grid with 12+log(O/H) = 7.53, the largest value of parameter S2
occurs at log(U) = --4.0, reaching the value of 0.37.
The nominal value of the observed parameter S2 = 0.31, midway between the values of S2
for the latter the lowest log(U) models, implies that the respective value of
its 12+log(O/H) falls midway between 7.35 and 7.53 dex, that is $\sim$7.44~dex.
For this combination, the expected ratio of F(3727)/F(H$\beta$) = 1.2 -- 1.4,
F(5007)/F(H$\beta$) $\lesssim$ 0.02.
It is worth to noting that despite the current data on the value of parameter
S2 allows us to safely assign the N region to the very low metallicity regime, the higher
S-to-N fluxes for the [S{\sc ii}] doublet are required to increase the accuracy of the
derived O/H.

It is also interesting to compare the spectra of the N knot in KK242
with that of the \HII\ region 'A' (a brighter of two known) in the nearby
dwarf galaxy Pegasus DIG (DDO~216=UGC~12613).
In some papers this galaxy is assigned to dTr due to its morphology and very low
current SFR. Its the adopted modern distance, derived with TRGB by \citet{Jacobs09},
is of 0.97~Mpc.  With the  total blue magnitude B$_{\rm tot}$ =13.22 and
M$_{\rm B}$ = --12.43  (HyperLEDA), PegDIG is almost 2 magnitudes more luminous than KK242.
The spectrum of this \HII\ region was analyzed by \citet{Skillman97}.
Its parameter S2 = 0.76$\pm$0.05 is $\sim$2.5 times larger than
the observed in the N knot of KK242 (0.31$\pm$0.15).
The lines visible in the DDO216 region 'A' spectrum in the range of 3700 -- 5300~\AA,
indicate very soft ionising radiation, with the flux ratio of [O{\sc iii}]$\lambda$5007
and [O{\sc ii}]$\lambda$3727 (hereafter parameter O32) less than 0.03
(at the 2$\sigma$ level). As the authors argue,
that low excitation spectrum can be explained  only by a sufficiently low
effective temperature of an exciting star, namely, T$_{\rm eff} \lesssim$ 32.5 kK,
consistent with a star B0V. As our {\sc Cloudy} grids evidence, this value of S2 indeed
emerges for the case of gas metallicity of 12+log(O/H)=7.92 as estimated by
\citet{Skillman97},  for a star with T$_{\rm eff}$ = 30~kK when log(U) = --4.0.
The respective parameter O32 appears of $\sim$0.015, consistent with the observed
upper limit. However, the model prediction for this case for the relative flux of
[O{\sc ii}]$\lambda$3727 and H$\beta$ appears $\sim$2.0, a factor of 1.7 lower
than the  observed one.

\subsubsection{Alternative value of KK242 gas metallicity and a possible related scenario}
\label{ssec:inflow_OH}

In the previous sections we presented the arguments that  in the  N knot of KK242,
the measured parameter S2 is well consistent with the value of the gas O/H,
expected from \citet{Berg12} reference relations for log(O/H) versus M$_{\rm B}$
and versus M${*}$, and also with the reduced gas metallicity as expected for
a void galaxy.  With this low gas metallicity
one could think on the typical late-type galaxy secular evolution and the 'recent' loss
of the main gas mass and the related drop of the 'normal' star formation.

Since the real uncertainty of the nominal value of S2 is $\sim$100\%,
we should check the variant of interpretation of a twice larger value of S2, that is 0.62,
or log(S2) = --0.21.
As one can see in Fig.~6, this value of log(S2) is close to the
{\sc Cloudy} model points for log(U) = --4.0 (red dotted line) at 12+log(O/H)
$\sim$ 7.8~dex that is too high for void dIrrs with the luminosity and mass similar
to that of KK242.

What kind of scenario could result at that elevated gas metallicity for the very low
stellar mass and luminosity of KK242?
One of the possible variants is related to the gas inflow to KK242 from
the outer parts of the disc of NGC6503 in course of their pericenter passage.
We did not find in the literature the published estimates of the gas
metallicity in NGC6503 despite its 2D spectra in the range 3600--6800~\AA\ were obtained
with the integral field unit  VIRUS-P \citep{VIRUS-P}.
Therefore, we adopt for NGC6503 the expected gas metallicity, that follows from the relation in
\citet{Berg12} for the Local Volume late-type galaxies. For its absolute magnitude of
M$_{\rm B}$ = --19.1~mag (HypeLEDA, averaged on several sources),
the expected value of 12+log(O/H) is $\sim$8.37 $\pm$ 0.15~dex. For  disc galaxies
with the visible metallicity gradients, this parameter is usually adopted at the radial
distance of r$_{\rm eff}$/2. Thus, taking into account rather small metallicity gradients
in the subluminous galaxies like NGC6503 ($\sim$0.02--0.03~dex~r$_{\rm eff}^{-1}$),
we expect the gas metallicity in the outer disc of NGC6503 at the level
of 12+log(O/H) $\gtrsim$ 8.1--8.2~dex.

Therefore, if we accept the hypothesis of gas inflow from
NGC6503 to the extremely gas-poor predecessor of dTr KK242, with the subsequent triggered
episode of star formation, we should expect the metallicity of gas in the N knot,
corresponding to 12+log(O/H) $\gtrsim$8.1$\pm$0.15~dex.
In Fig.~6, the 12+log(O/H) = 8.1~dex for {\sc Cloudy} models with log(U) = --4.0
(red diamonds) corresponds
to the value of S2 $\sim$1.0 (log(S2) = 0), that is more than 2$\sigma$ larger
than the nominal value  S2 = 0.31.

These estimates adopt the gas metallicity of NGC6503 based on the relation
established by \citet{Berg12} on the subsample of the late-type galaxies within
the Local Volume,
and hence, should be nicely applicable to NGC6503. However, as discussed in our papers
\citep{PaperVII, XMP-SALT, XMP-BTA}, this sample is mostly related to the typical
groups and their environs. For galaxies with the same M$_{\rm B}$ residing in the nearby
voids, the gas metallicity (or log(O/H)) is in average lower by 0.14~dex.
For 'luminous' galaxies similar to NGC6503, this effect is smaller,
$\sim$0.1~dex \citep[][Fig.1]{XMP-BTA}.
Since NGC6503, as described in Section~\ref{ssec:environs}, resides in the void Dra--Cep,
we expect, that the above estimates of 12+log(O/H) should be reduced.
That is the expected value of gas O/H in the N knot, in the framework of the hypothesis
with the gas inflow from the outer disc of NGC6503, is 12+log(O/H) $\gtrsim$ 8.0~dex.

We summarize this attempt to relate the highest possible value of S2 (about two S2 nominal)
and the respective value of 12+log(O/H) $\sim$7.8~dex for the KK242 N knot
with the 'metal-rich' gas from the outer parts of NGC6503 (12+log(O/H) $\gtrsim$8.0~dex),
as follows. The gap between the two extreme possible values for the gas metallicity in KK242 and
NGC6503 remains too large. So that it is hard to reconcile this hypothesis with the available data.

\subsection{Fading stages of faint SF episodes and the problem
of gas metallicity estimate}
\label{ssec:OH_fading}

In the light of the discussed above dwarf galaxies with only the weak tracers
of the recent SF episode, we would like to emphasize that this type of objects
should be numerous and widely spread among the low mass dwarfs. Due to their
low masses, their gas metallicity is expected to be in the low-Z regime.

Indeed, in a sizeable fraction of low mass late-type LSB galaxies, the observed SFR is
subtle. See, e.g., \citet{Kaisin2019} for dwarfs in the Local Volume and
\citet{SHIELD.OH} for small gas-rich dwarfs selected from the ALFALFA
survey \citep{ALFALFA18} and IZw18C \citep{Izotov2001}.
The observed H$\alpha$ luminosities per individual \HII\ region indicate a small
number of the hot massive stars in such dwarfs. The LV dwarfs with the lowest observed
luminosities of L(H$\alpha$) of the order of $\lesssim$10$^{36}$ erg~s$^{-1}$
correspond to the ionising photon fluxes Q$_{0}$ of stars O9V and later.
Several XMP galaxies from our recent papers \citep{XMP-SALT,XMP-BTA} fall to
this regime as well. In addition, the new current and upcoming deep sky surveys
will drastically increase the number of such galaxies.

For an instantaneous SF episode, the probability
to catch it in the early phases (say, younger than $\lesssim$ 8--10 Myr) when
sufficiently hot stars (conditionally, O5V-O8V) are
still alive\footnote{if they were really formed in the modest total mass involved.}
and provide a high enough log(U) and T$_{\rm eff}$, is lower than to
catch a region with the exciting stars of O9V-B0V,
with the lower fluxes of the ionising photons and  T$_{\rm eff}$.
In such cases, if we wish to
obtain the estimates of gas metallicity, we need some alternative means.

For that low excitation conditions and the related very low fluxes of
[O{\sc iii}] lines, the possibility to use the 'strong-line' empirical methods
to estimate their gas metallicity is hampered. Due to various observational
selection effects, such objects remain largely underexplored, in particular,
in the context of their gas metallicity.

We suggest to use for such low excitation \HII\ regions with the Hydrogen Balmer
series lines and only a few heavy element lines detected -- [O{\sc ii}]$\lambda$3727,
 [S{\sc ii}]$\lambda$6716,6731 doublets and probably [N{\sc ii}]$\lambda$6584, the
grid of {\sc Cloudy} package models with the low log(U) and low T$_{\rm eff}$ as typical
for \HII\ regions ionized by the early B-type and late O-type stars.

The examples of such analysis for KK242 and DDO216 presented above,
show that this is feasible. The main pre-requisite of the successful application of
such models is a sufficiently good S-to-N ratio for the used heavy element line fluxes.

The used here the {\sc Cloudy} grids are based on the models of the central
star with T$_{\rm eff}$. The more advanced grids, with the modern stellar atmosphere
models and the stellar metallicity included,
can give us an advanced mean to treat spectra and derive a more reliable estimates of
the gas metallicity in the  low excitation \HII\ regions in a large number of dwarf
galaxies within the Local Volume and its environs. This, in turn, should allow one to address
the issue of chemical evolution on a wider range of galaxy parameters.

The output of such model grids in the form of the relative line fluxes can be used
similar to the 'Counterpart' method of \citet{Counterpart2012} which seeks in
the reference database of the observed \HII\ regions with known direct O/H
for a combination of the relative line fluxes which is the closest to that
in the studied \HII\ region without detected [O{\sc iii}]$\lambda$4363 line.

\subsection{Physical parameters and star formation}
\label{ssec:SF}

The dTr galaxy KK242 is very interesting in the context of its current and
recent star formation. Its gas mass is several times smaller than the typical
of the comparable stellar mass and luminosity late-type LSB dwarfs. Therefore,
if its \HI\ gas distribution is not strongly clumped, this suggests the reduced
column density. Indeed, as the VLA map in the \HI-line reveals \citep{KK242.HST},
the peak \HI\ column density in this galaxy reaches only 3$\times$10$^{19}$~atom~cm$^{-2}$.
This is to compare with the typical threshold gas column density for the onset of
star formation in dwarf irregulars, blue compact and LSB galaxies defined at
the level of 1.0$\times$10$^{21}$~atom~cm$^{-2}$ for the linear scales of $\sim$0.5~kpc
\citep{skillman1987, taylor94, Begum08, Ekta08}.
This can be partly explained by the large effective beam-size of $\sim60 \times 40$ arcsec$^2$
(or 1.85~$\times$ 1.24~kpc), which smears the higher density features at smaller
linear scales. However, the significant column gas underdensity of the KK242 SF
region attracts the special attention to this point.

The issue of the threshold column density for the onset of SF is not
settled, however \citep[see, e.g., discussion in][]{Ekta08}. The model
calculations \citep[e.g.][]{Schaye2004} predict a range for this parameter which depends
on gas mass fraction, pressure, its metallicity, ionising flux radiation.
Therefore, it will take much effort to understand whether the formation of the studied complex
of several massive hot stars in KK242 took place in a low density gas under the special
conditions.

The important factor of the SF episode onset is the proximity of KK242 to its host NGC6503.
As many N-body simulations indicate \citep[e.g.][]{diMatteo}, the peak of the tidally induced
SF episode occurs in a few hundred Myr after the first pericenter passage. We can roughly estimate
the time since the passage of KK242 near the host. Taking the relative tangential component
of velocity ($\delta V_{\rm tang}$) approximately equal to the relative radial velocity
$\delta V_{\rm rad} \sim$100~\kms, and the mutual projected distance of $\delta r \sim$ 31~kpc,
we estimate the respective time t$_{\rm passage} \sim$ 300~Myr. Therefore, the presence in the
discussed SF region of RHeB and BHeB stars, with ages of $\sim$100~Myr, does not contradict
to the assumption that the long-lasting very localised SF episode was
triggered by the strong tidal interaction with the massive host. To get a deeper insight into
gas properties involved in the SF episode at such atypical conditions, it seems, one needs to wait for
the ngVLA, which will allow one to obtain \HI\ data simultaneously with
the high sensitivity and suitable angular resolution.

\section[]{Summary and conclusions}
\label{sec:summary}

In this study we primarily interested in the determination of the radial
velocity of the ionised gas in KK242 in order to get its value independent
on the previous \HI\ observation. Thanks to the proper observational conditions, we
spatially resolved two faint emission regions (the N and S knots) within the studied
SF complex. Their appearance is rather different that pushed us to analyze
them individually. Thanks to the publicly available F606W and F814W HST images of
KK242, as well as of the photometry of its individual stars, we were able
to disentangle the exciting stars of the observed nebular emission.
In the S knot we also identify a nebula, a likely SN remnant.

Due to rather low S-to-N of the available spectra of this SF complex and comparatively
low effective temperatures of the ionising  massive stars, the only heavy
element lines detected so far is the [S{\sc ii}]$\lambda\lambda$6716,6731 doublet.
Its flux ratio to that of H$\beta$ (parameter S2) can be used as a rough empirical
indicator of the gas metallicity. However, the S2 in our data has the large
uncertainty that allows rather wide range of O/H. If we compare
the observed S2 in the N knot of KK242 with the data for the sample of \HII\ regions
with the directly derived O/H (that is with the medium or high excitation), it corresponds
to values of 12+log(O/H) between 7.6 and 8.0~dex.

However, the examination of our {\sc Cloudy} package grids with the wide range of
lg(U) and T$_{\rm eff}$ of ionising stars, reveals the elevated values of S2 for
the 'extremely' low values of lg(U) $\sim$ --4.0 and T$_{\rm eff}$ $\sim$ 25--30~kK.
These elevated values of S2 are consistent with the observed one
for 12+log(O/H) as low as $\sim$7.45~dex.  Meanwhile, the mentioned above
'extremely' low lg(U) and T$_{\rm eff}$ are consistent with those expected
for B0V stars directly observed in this region at the HST images.
The possibility of that 'low' value of O/H is important, since this is
expected for KK242 luminosity and stellar
mass from the reference relations for the LV late-type galaxies in \citet{Berg12}.
The gas metallicity is a crucial parameter for the choice between possible evolutionary
scenarios. The currently available estimate of O/H 
is consistent with  the case
of the typical secular chemical evolution of late-type dwarfs as possible predecessors
of the dTr KK242. This conclusion remains valid if we take into account
that the evolution of KK242 took place in a void, and therefore one expects it
to be reduced.

We also examine an alternative scenario involving the higher metallicity
gas inflow from
the outer disc of NGC6503 to the 'gas-free' dE predecessor of KK242
after its pericenter passage. From the estimates of the possible metallicity
of the in-flowed gas, this variant seems to be improbable due to the substantial
gap between the upper limit of the gas metallicity in KK242 and the lower limit of that in
the outer part of NGC6503.

Therefore, the most likely scenario of the origin of KK242 as a void dTr, combines
its secular evolution as a void low-mass dIrr and the 'recent' rapprochement and interaction
with the much more massive host NGC6503. The  pericenter passage of KK22 several
hundred Myr ago resulted in the tidal stripping of its gas and triggered the episode of
SF \citep[e.g.,][]{DiCintio2021}. The traces of this 'recent' star formation are seen
in the GALEX UV images as light of stars with the ages of less than a few hundred Myr
\citep{Koda2015} as well as BHeB and RHeB stars in the studied here region with the
faint H$\alpha$ emission.

Summarising all available data and the above analysis and discussion, we arrive
at the following conclusions.

\begin{enumerate}
\item
KK242 is a transition type dwarf residing in a nearby void. We report its
BTA spectroscopy and the new value of its radial velocity,
V$_{\rm hel}$ = --66$\pm$10~\kms,  based on the observed H$\alpha$ line in
two adjacent regions of the star-forming complex identified by \citet{Koda2015}.
This value is consistent with that of the recently found
faint \HI\ emission from KK242 \citep{KK242.HST} and is lower
than the radial velocity of its host spiral NGC6503 by $\sim$100~\kms.
\item
The appearance of these two regions in the BTA 2d spectra of KK242
looks very different due to the bright light of the 'unrelated'
Red Supergiant at the projected angular distance of $\sim$0.7~arcsec from the
H$\alpha$ intensity peak of the Northern H$\alpha$ knot.
On the HST images of KK242, we identify the sources responsible for the visible
H$\alpha$ emission within the BTA long slit.
The Northern region is a superposition of three \HII\ regions exciting by
the late O-type and early B-type stars at the mutual projected distances of
$\sim$0.7--1.0~arcsec (20--30 pc). The H$\alpha$ emission of the
Southern region is a superposition of an \HII\ region excited
by an early B-star and of a probable supernova remnant at a projected
distance of $\sim$0.7~arcsec.
\item
Besides the early-type blue stars, we identify within the boundaries
of this SF complex three additional luminous stars. They are tentatively
classified as one RHeB star (the mentioned above Red Supergiant) and two
BHeB stars.
Their colours and absolute magnitudes imply their ages of $\lesssim$100~Myr.
This suggests that the recent SF episode as traced by several massive
young main-sequence stars (O9-B1) in fact lasts in this location
at least of  $\sim$0.1~Gyr or so.
\item
Due to rather low S-to-N ratio spectrum of the Northern
\HII\ region and its low excitation, the only detectable metal lines
appear those of the [S{\sc ii}] doublet. We use the parameter S2 (the flux ratio
of [S{\sc ii}] to that of H$\beta$) to constrain the gas metallicity in this region.
The direct comparison of the nominal value of S2(KK242,N) = 0.31 with the \HII\ regions from the
compilation of \citet{Counterpart2012}, allows the wide range of 12+log(O/H) = 7.6--8.0~dex.
This range, however, is poorly consistent with the low 12+log(O/H) $\sim$7.4$\pm$0.1~dex,
expected for KK242 low stellar mass and luminosity in the case it is treated as a dIrr
which lost most of its gas.
\item
We undertake a deeper analysis of this situation taking into account the
HST data on the ionising stars' parameters (O9V--B0V). We use the {\sc Cloudy} package
to construct  grids of the common line flux ratios in \HII\ regions with the
wide range of gas metallicities, ionisation parameter U and the effective temperature
of the central star. The observed parameter S2 in the Northern knot of KK242 can be
 well consistent with the predicted in models with 12+log(O/H) $\sim$7.4~dex,
if both log(U) and T$_{\rm eff}$ are very low, $\sim$-4.0 and $\sim$25--30~kK, respectively.
These log(U) and T$_{\rm eff}$ are close to those expected for the massive hot stars visible
in this region.
\item
For the Southern H$\alpha$ knot, the higher S-to-N value of flux of
the [S{\sc ii}] doublet indicates its elevated ratio relative to the
flux of H$\alpha$. We argue that since the main contribution to the
emission of this knot comes from the nearby nebula, a likely SN remnant,
this elevated [S{\sc ii}] emission is related to the shock excitation in the
SNR shell.
\item
We pay attention to the generalisation of the problem of gas metallicity
determination in common low-excitation low-metallicity \HII\ regions of LSB dwarfs.
In such regions, only a few heavy element emission lines are typically observed and the use
of the popular strong-line empirical estimators can be impossible.
We suggest to develop a grid of {\sc Cloudy} package models representing the
observed line ratios for the wide range of gas and young star metallicity
when only a population with the ages of more than 10--15 Myr excites their related \HII\
regions (late O and early B-type). This will give one a new advanced mean to
address the gas low metallicity in the dwarfs of the Local Universe with the
low/subtle current SF.
\end{enumerate}

\section*{Acknowledgements}
The work was supported by the Russian Scientific Fund (RScF) grant
No.~22-22-00654.
The authors thank I.D.~Karachentsev for sharing some results on KK242
before publication. We also thank D.I.~Makarov
for the help with the use of the publicly available HST data.
We acknowledge the constructive suggestions of the anonymous
referee which helped us to improve and clear up  the paper contents.
The authors acknowledge the allocation of the SAO DDT time at BTA
in November 2021.
Observations with the SAO RAS telescopes are supported by the Ministry of Science and
Higher Education of the Russian Federation. The renovation of telescope equipment is
currently provided within the national project "Science and Universities".
This research is partly based on observations made with the NASA/ESA Hubble
Space Telescope obtained from the Space Telescope Science Institute,
which is operated by the Association of Universities for Research in
Astronomy, Inc., under NASA contract NAS 5-26555. These observations are associated
with program SNAP-15922.
We acknowledge the use of the {\sc Cloudy} photoionisation code to model the intensities
of the common emission lines in \HII regions of the low excitation.
We acknowledge the use for this work of the database HyperLEDA\footnote{http://leda.univ-lyon1.fr}.
This research has made use of the NASA/IPAC Extragalactic Database (NED)
which is operated by the Jet Propulsion Laboratory, California Institute
of Technology, under contract with the National Aeronautics and Space Administration.

\section*{Data availability}
The data underlying this article are available
in the The Extragalactic Distance Database (EDD)\footnote{http://edd.ifa.hawaii.
edu/}.
The HST/ACS data used in this article are available in the
STScI data archive.


\bsp

\appendix

\section{[S{\sc ii}]$\lambda\lambda$6716,6731 doublet: the range of its strength
  via {\sc Cloudy}}
\label{sec:S2-variance}

In this Appendix we use the {\sc Cloudy.17} version of the {\sc Cloudy}
radiative transfer code, as described by \citet{Cloudy.17},
to perform calculations of the possible range for the value of parameter S2
(flux ratio of the doublet [S{\sc ii}]$\lambda\lambda$6716,6731 to that of H$\beta$)
for a range of \HII\ region gas metallicities. We took Oxygen abundances of 12+log(O/H)
= 7.00, 7.17, 7.35, 7.53, 7.70, 7.90  and 8.04~dex.

The relative abundances C/O, N/O, Ne/O, Si/O, S/O, Ar/O, Fe/O are adopted
as the averages found in the systematic study of 54 supergiant \HII\ regions with
the range of 12+log(O/H) = 7.1 -- 8.2~dex in the paper by \citet{Izotov99}.
Since the abundance of O/H is set in the units of the Solar one, we adopt 12+log(O/H)\sunn
= 8.69~dex according to \citet{Asplund2009}.

We construct a grid of models with the range of the ionisation parameter U (log(U) from -4.0 to -1.0,
with the step of 0.5, and the effective temperature of the central star (black-body) T$_{\rm eff}$
from 25 to 50 kK, with the step of 5~kK. We then apply these results to the
observational data for the Northern knot in KK242 (Fig.~6 and Section~\ref{ssec:OH_gas})
and attempt to constrain its metallicity from the observed parameter S2.

We run {\sc Cloudy} in the mode "sphere"  with the adopted input parameters
of the inner radius of R$_{\mathrm 0}$ = 10$^{17}$~cm and the number density of
the hydrogen nuclei of n(H)=10 cm$^{-3}$. {\sc Cloudy} then computes the
structure of the photoionised region by requiring that the density be constant.
The outer radius is assumed to be the one where the gas temperature falls
to 4 kK since the colder gas practically does not produce optical emission lines.
The resulting geometry of all models was  closed, that is, the gas covers most of
the central ionising source. The iterations were carried out until the optical depth of the line
and continuum became stable. For the gas element abundances, we do not take
into account a presumably small fraction of material which is possibly locked into grains.
We remark for clarity, that the ionisation parameter U shown on
the X-axes of Figs.~A1 and A2, is obtained as a result of a model and relates to the
Str\"omgren radius, in difference with an input parameter of a model at the inner radius.
That is, it directly corresponds to the parameter U, estimated for a B0V star in
Section~\ref{ssec:star_parameters}.

\begin{figure}
\includegraphics[width=7.0cm,angle=-90,clip=]{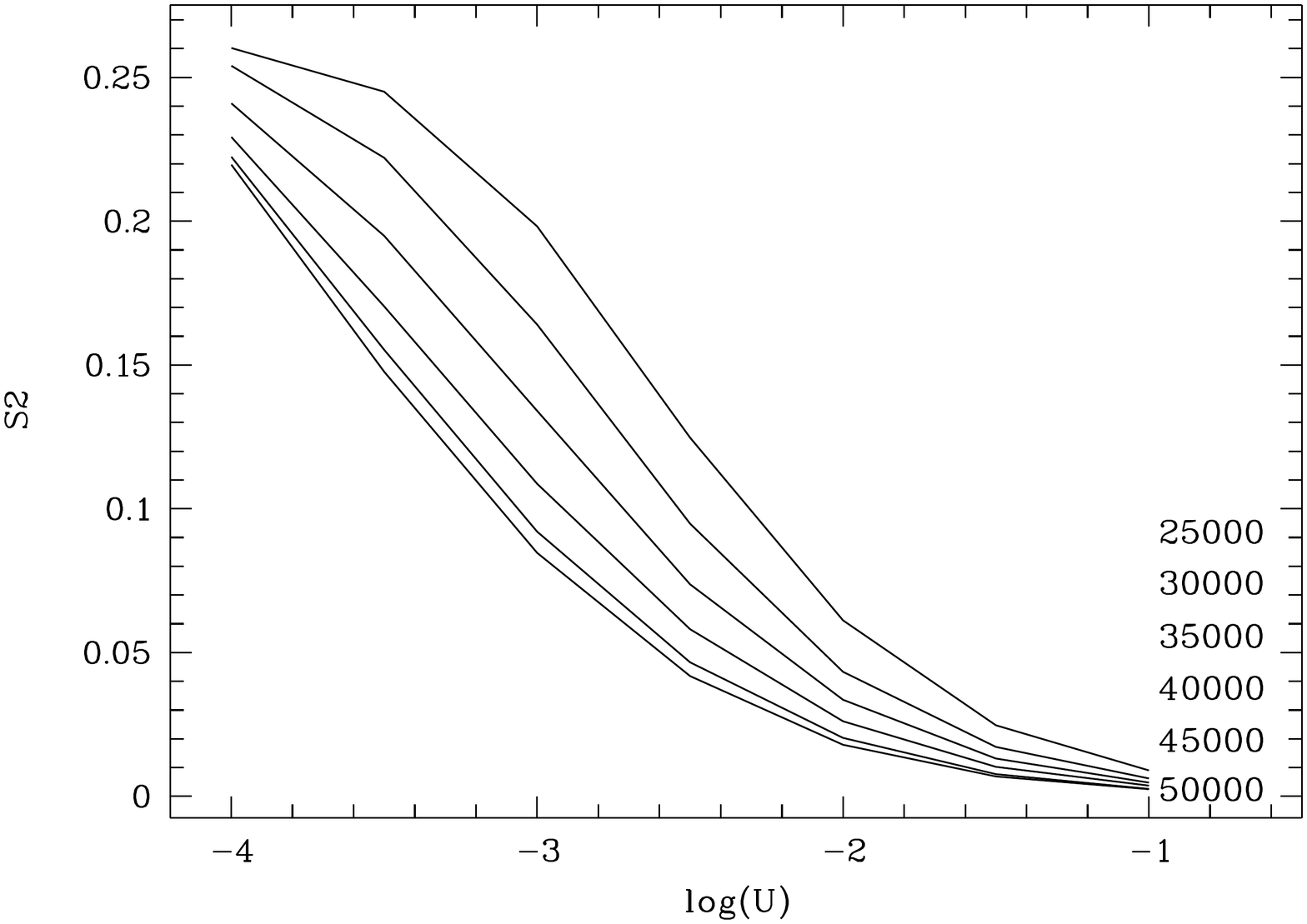}
\includegraphics[width=7.0cm,angle=-90,clip=]{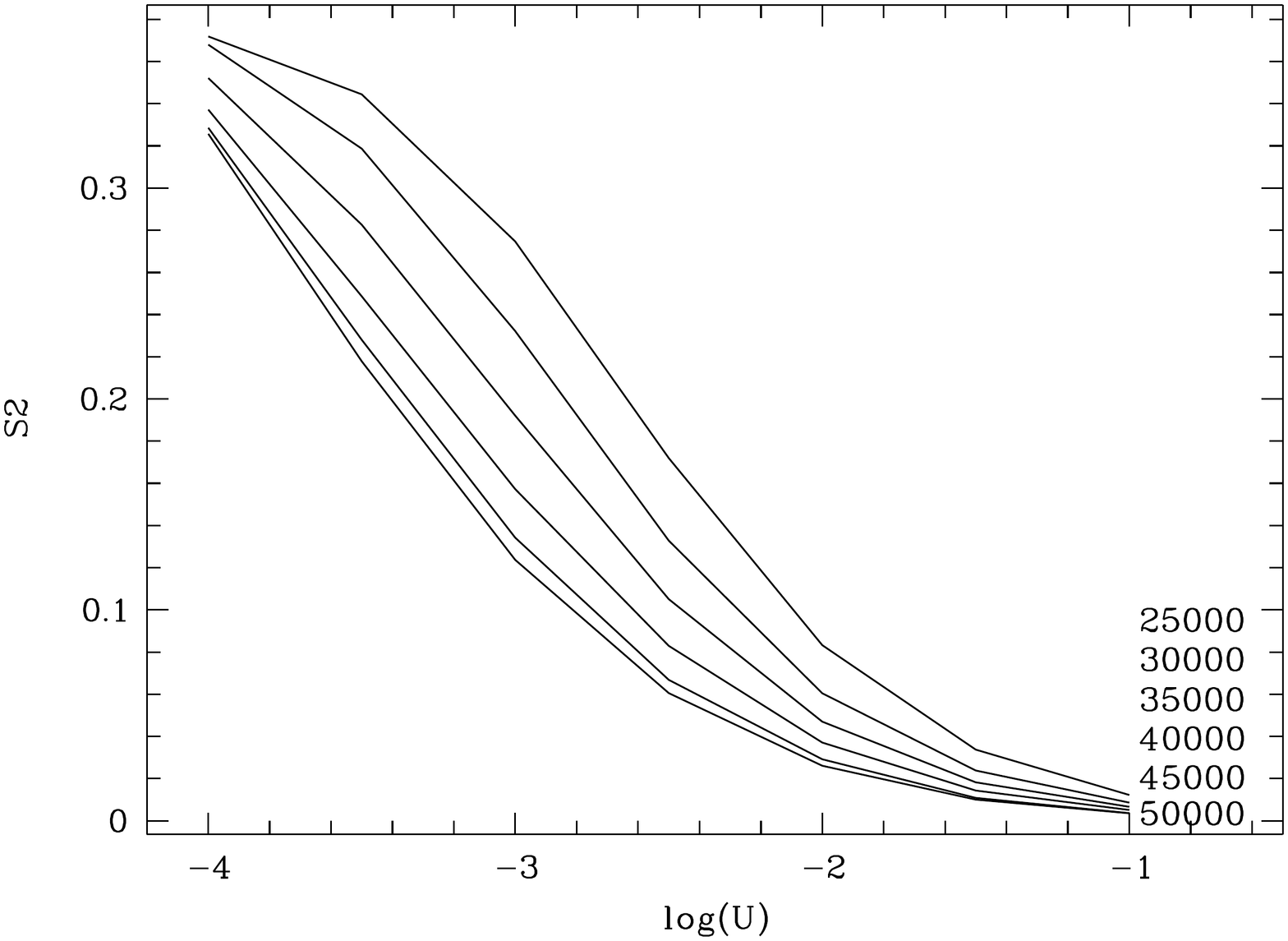}
\includegraphics[width=7.0cm,angle=-90,clip=]{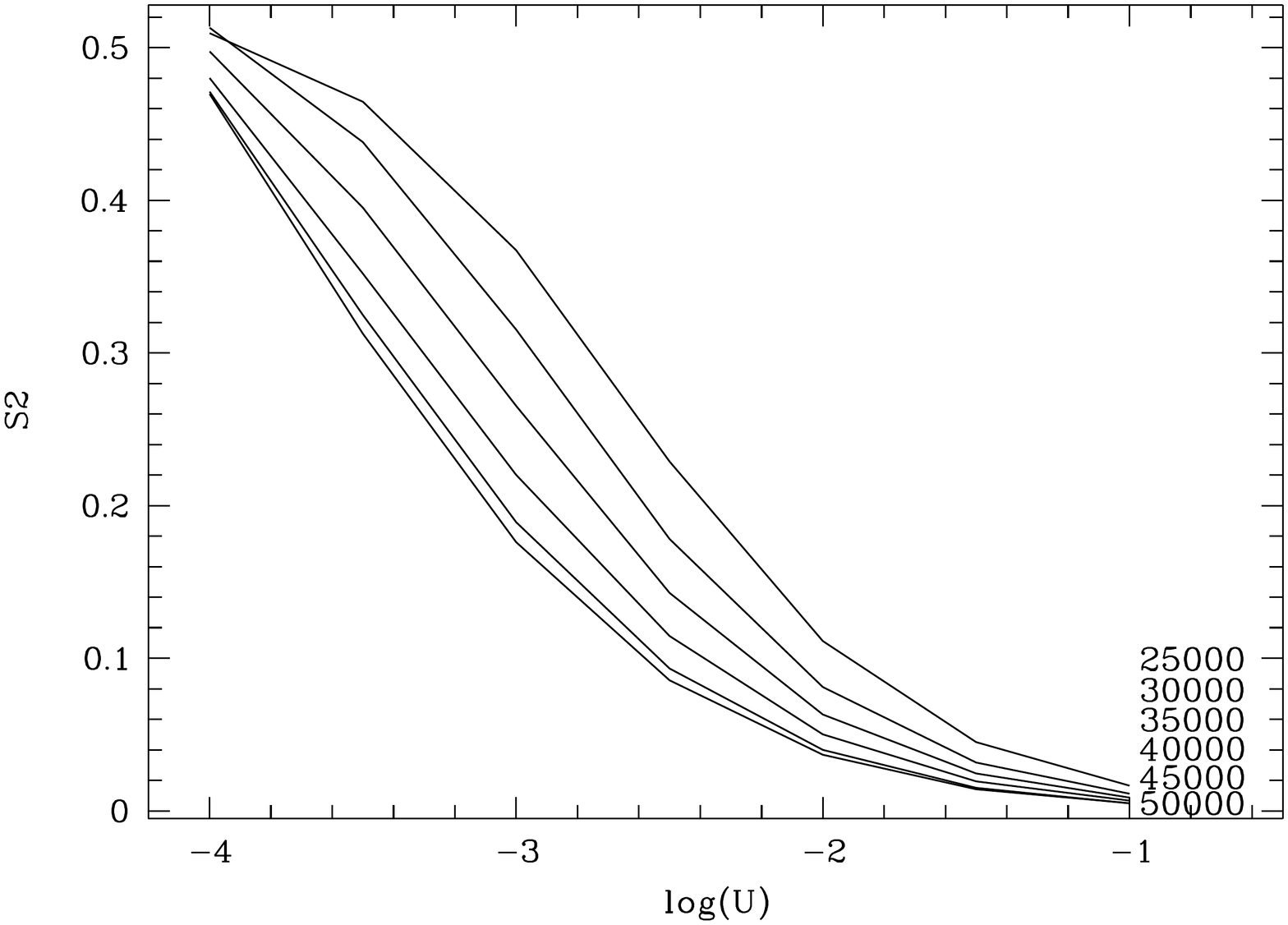}
\caption{
Plots showing variation of parameter S2 in an \HII\ region as a function of the
ionisation parameter log(U) for ionising central stars with the range of T$_{\rm eff}$
from 25 to 50~kK for three values of gas O/H, 12+log(O/H) = 7.35, 7.53 and 7.70,
from top to bottom, respectively, as obtained with the package {\sc Cloudy}.
For the fixed O/H, the S2 parameter appears a factor of 2 or so larger at
log(U) $\sim$ --4.0
than for \HII\ regions with  log(U) $\gtrsim$ --2.5, typical of the observed for
O/H (dir) (see Fig.~\ref{fig:Cloudy-4363}).
}
\label{fig:Cloudy-S2}
\end{figure}

\begin{figure}
\includegraphics[width=7.0cm,angle=-90,clip=]{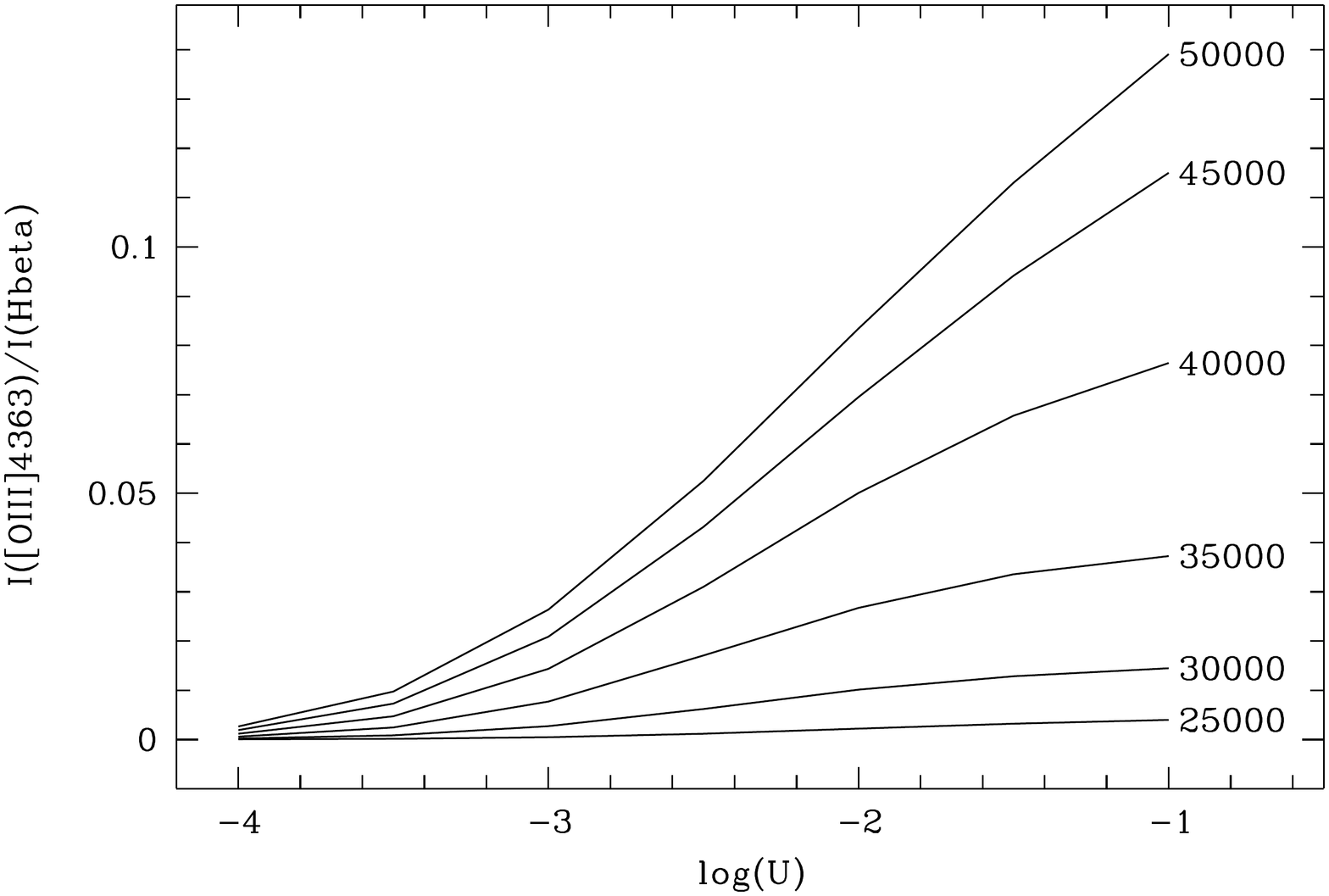}
\includegraphics[width=7.0cm,angle=-90,clip=]{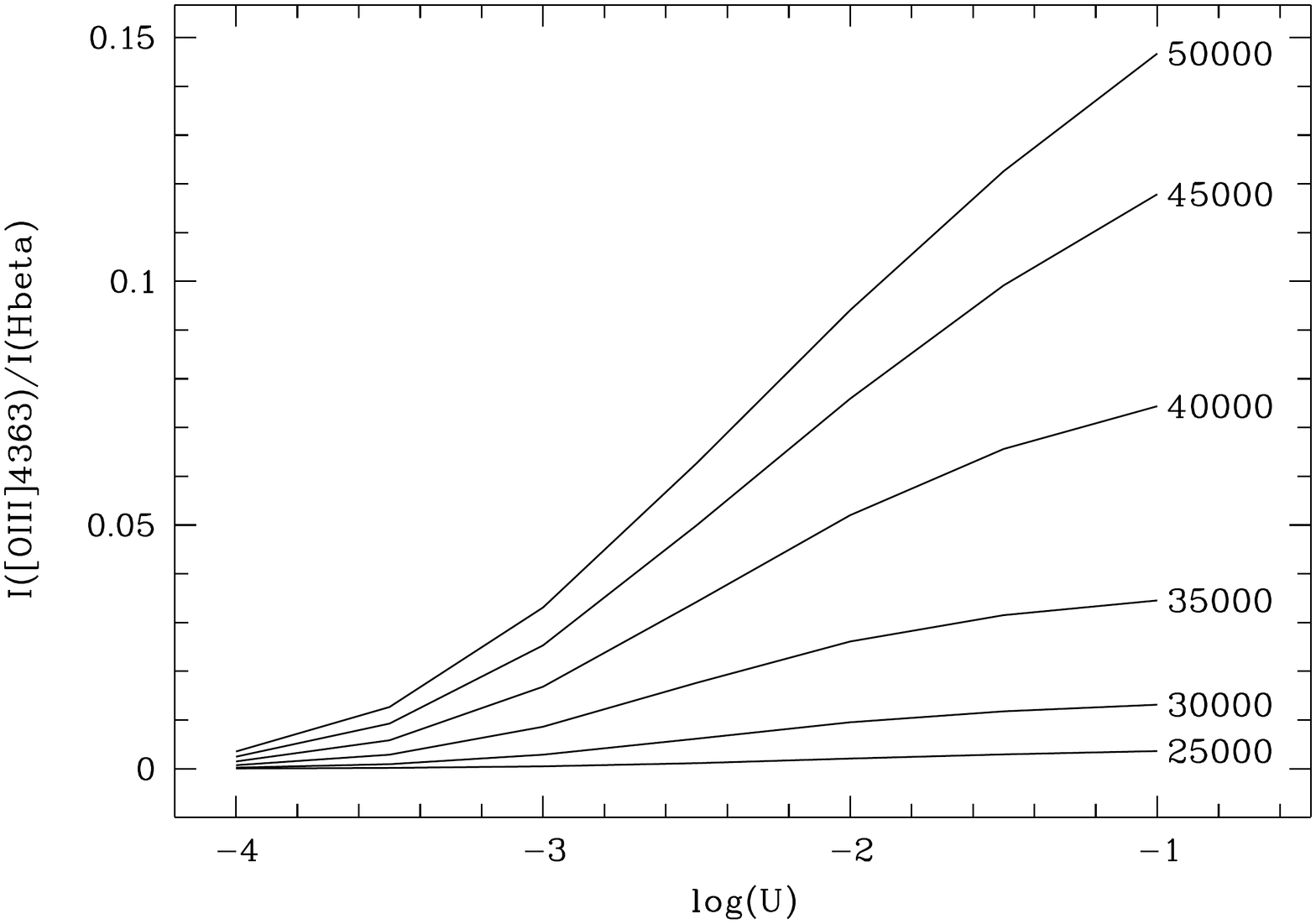}
\includegraphics[width=7.0cm,angle=-90,clip=]{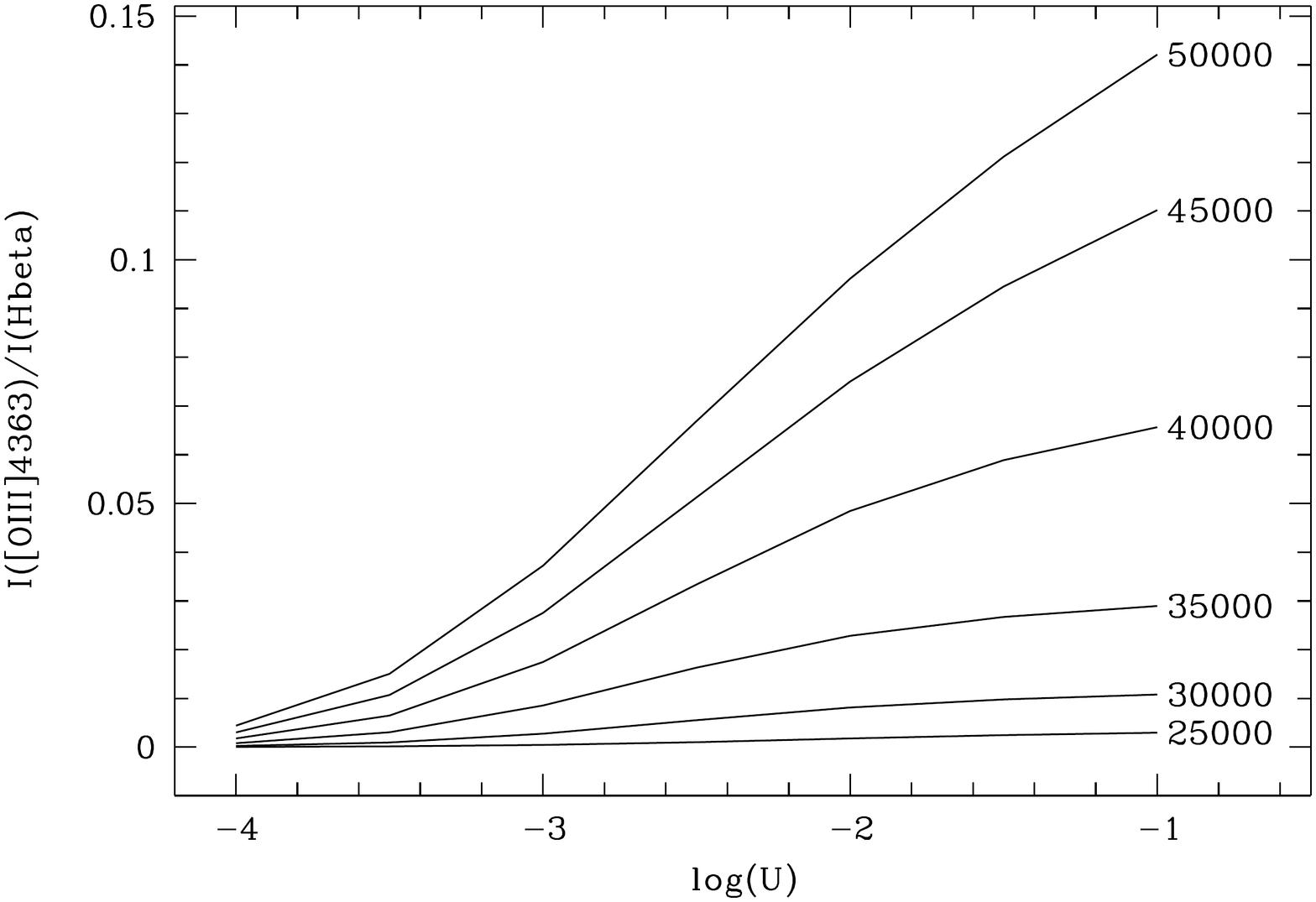}
\caption{
Similar plots as in Fig.~\ref{fig:Cloudy-S2} showing variation of the line flux ratio
[O{\sc iii}]$\lambda$4363 to H$\beta$.
For T$_{\rm eff} \lesssim$ 35 kK, this ratio exceeds the conditional level of 0.01
(for \HII\ regions with the direct O/H determination)  only for
models with the ionisation parameter log(U) $\gtrsim -3.0$.
}
\label{fig:Cloudy-4363}
\end{figure}

\label{lastpage}

\end{document}